\documentclass[journal]{IEEEtran}
\usepackage{amsmath,amsfonts}
\usepackage{algorithmic}
\usepackage{algorithm}
\usepackage{array}
\usepackage[font=small,labelfont=bf]{caption}
\usepackage{textcomp}
\usepackage{stfloats}
\usepackage{url}
\usepackage{verbatim}
\usepackage{graphicx}
\usepackage{pgfplots}
\usepackage{cite}
\usepackage{multirow}
\usepackage[colorlinks,citecolor=blue]{hyperref}
\usepackage{tikz}

\usepackage{array}
\newcolumntype{P}[1]{>{\centering\arraybackslash}p{#1}}

\renewcommand{\arraystretch}{1.05}

\makeatletter
\newcommand{\pushright}[1]{\ifmeasuring@#1\else\omit\hfill$\displaystyle#1$\fi\ignorespaces}
\newcommand{\pushleft}[1]{\ifmeasuring@#1\else\omit$\displaystyle#1$\hfill\fi\ignorespaces}
\makeatother

\input epsf
\usepackage{graphicx}
\usepackage{url}

\DeclareRobustCommand*{\IEEEauthorrefmark}[1]{%
  \raisebox{0pt}[0pt][0pt]{\textsuperscript{\footnotesize\ensuremath{#1}}}}

\begin{document}

\title{\LARGE \textsc{Designing optimal loop, saddle, and ellipse-based magnetic coils by spherical harmonic mapping}}
\author{\IEEEauthorblockN{Peter~James~Hobson\IEEEauthorrefmark{1,\text{a},\text{b}},
Noah~Louis~Hardwicke\IEEEauthorrefmark{1,\text{b}},
Alister~Davis\IEEEauthorrefmark{1},
Thomas~Smith\IEEEauthorrefmark{1},
Chris~Morley\IEEEauthorrefmark{1},
Michael~Packer\IEEEauthorrefmark{1},
Niall~Holmes\IEEEauthorrefmark{1,2},
Max~Alain~Weil\IEEEauthorrefmark{1},
Matthew~Brookes\IEEEauthorrefmark{1,2},
Richard~Bowtell\IEEEauthorrefmark{1,2}, and
Mark~Fromhold\IEEEauthorrefmark{1}} \\
\IEEEauthorblockA{\IEEEauthorrefmark{1}School of Physics and Astronomy, University of Nottingham, NG7 2RD, United Kingdom} \\
\IEEEauthorblockA{\IEEEauthorrefmark{2}Sir Peter Mansfield Imaging Centre, University of Nottingham, NG7 2RD, United Kingdom} \\
\IEEEauthorblockA{\IEEEauthorrefmark{\text{a}}\href{mailto:peter.hobson@nottingham.co.uk}
{peter.hobson@nottingham.ac.uk}} \\
\IEEEauthorblockA{\IEEEauthorrefmark{\text{b}}These authors contributed equally to this work} \\}


\maketitle

\begin{abstract}
Adaptable, low-cost, coils designed by carefully selecting the arrangements and geometries of simple primitive units are used to generate magnetic fields for diverse applications. These extend from magnetic resonance and fundamental physics experiments to active shielding of quantum devices including magnetometers, interferometers, clocks, and computers. However, finding optimal arrangements and geometries of multiple primitive structures is time-intensive and it is challenging to account for additional constraints, e.g. optical access, during the design process. Here, we demonstrate a general method to find these optimal arrangements. We encode specific symmetries into sets of loops, saddles, and cylindrical ellipses and then solve exactly for the magnetic field harmonics generated by each set. By combining these analytic solutions using computer algebra, we can use numerical techniques to efficiently map the landscape of parameters and geometries which the coils must satisfy. Sets of solutions may be found which generate desired target fields accurately while accounting for complexity and size restrictions. We demonstrate this approach by employing simple configurations of loops, saddles, and cylindrical ellipses to design target linear field gradients and compare their performance with designs obtained using conventional methods. A case study is presented where three optimized arrangements of loops, designed to generate a uniform axial field, a linear axial field gradient, and a quadratic axial field gradient, respectively, are hand-wound around a low-cost, 3D-printed coil former. These coils are used to null the magnetic background in a typical laboratory environment, reducing the magnitude of the axial field along the central half of the former's axis from $\mathbf{\left(7.8\pm0.3\right)}$~$\boldsymbol{\mu}$T (mean $\mathbf{\pm}$ st.~dev.) to $\mathbf{\left(0.11\pm0.04\right)}$~$\boldsymbol{\mu}$T.

\end{abstract}


\bstctlcite{IEEEexample:BSTcontrol}

\section{Introduction}
Evermore sophisticated methods of magnetic field design are required to better null unwanted variations and generate targeted biases in precision measurement devices. Techniques to control magnetic fields have been applied for various precision measurements including in magnetic resonance scanners~\cite{doi.org/10.1002/cmr.a.20163,10.1016/j.jmr.2016.06.015,
9969569}, electric dipole moment experiments~\cite{PEREZGALVAN2011147,10.1063/1.4894158,doi:10.1119/1.5042244}, and Kibble balances for mass determination~\cite{Sutton_2014,9042232,Schlamminger_2022}. As well as this, precision magnetic field control is a necessity for new generations of quantum-enabled devices, including atomic magnetometers~\cite{HOLMES2018760,PhysRevApplied.15.054004,PhysRevApplied.15.064006,9743468,app12168219,doi:10.1063/5.0102402}, atom interferometers~\cite{Ji:2021qhl,Hobson_2022}, atom~\cite{10.1016/j.actaastro.2014.06.007, Petrov_2018} and ion~\cite{doi:10.1063/5.0049734} clocks, and quantum computers~\cite{goldman2000magnet,doi:10.1063/1.4966970}. Recently, the design of on-board coils for quantum-enabled devices has garnered extensive research interest~\cite{doi:10.1063/1.5036605,8731911,9873890,PhysRevApplied.18.014036} as these coils must balance magnetic constraints, such as power-efficiency and field uniformity, with significant non-magnetic considerations, including optical access and ease-of-assembly.

Turner~\cite{10.1088/0022-3727/19/8/001} pioneered the design of target coils from discretized surface currents, in which wires emulate a continuum of current flowing on a surface. Typically, the surface current is decomposed into a set of weighted orthogonal modes~\cite{10.1088/0022-3727/34/24/305,10.1088/0022-3727/35/9/303,10.1088/0022-3727/36/2/302}. As these modes generate spatially orthogonal magnetic fields, the best combinations of weightings to generate a specified target field profile may be determined using simple mathematical programming~\cite{https://doi.org/10.1002/mrm.1910260202}, often least-squares minimization. However, the real-world performance of surface current-based coil systems is limited if the surface current is inaccurately emulated by the wire pattern. This is known as \emph{discretization error}~\cite{DiscretePaper}. This error increases if the target field region is close to the coil or if manufacturing limitations prevent precise emulation of the surface current. In some cases discretization error may be calculated analytically~\cite{doi:10.1063/5.0063054,10.1016/j.nima.2013.05.013}, but often it may only be ascertained by numerically simulating each wire pattern \emph{a posteriori} until a well-represented design is found. This may be very computationally intensive.

Unlike surface current-based coils, those directly designed from simple building blocks, including loops, saddles, and ellipses, hereby referred to as \textit{primitives}, do not suffer from discretization error and have regular shapes, making them cheap and easy to manufacture. Romeo~and~Hoult~\cite{Romeo} developed a method to design magnetic fields using these coils by exploiting the symmetries in a desired magnetic field, expressed using a spherical harmonic decomposition, with sets of primitives. They then maximized the target harmonic strength by tuning both the continuous geometric \emph{coil parameters}, which determine the shape of the sets of primitives, and the discrete ratios of the number of turns of wire among different sets of primitives. Arrangements of multiple sets of primitives have the capability to produce a high quality magnetic field since each extra primitive provides additional parameters that one can optimize. However, the optimization of multiple primitives is challenging as each primitive generates many spherical harmonics. To design such systems, one may examine Taylor expansion coefficients~\cite{10.1063/1.1716184,10.1002/cmr.b.20057,9815315,9481934} of analytic expressions~\cite{Simpson2003,doi:10.1063/1.5036605}, but these do not relate simply to the basis of spherical harmonics in many cases. This can make it difficult to ascertain how the minimization of different undesired variations should be prioritized to maximize field quality. Alternatively, numerically-solved analytic formulations of the magnetic field~\cite{osti_4156720,9612088,https://doi.org/10.1002/bem.2250130507} have been used effectively to design loops and saddles, but such approaches do not allow the relationship between the coil parameters and field quality to be determined \emph{a priori}. As a result of these limitations, surface current-based coils are often preferred over primitive-based systems in settings where accurate target fields are required over large target regions relative to the coil size.

Here, we facilitate the wider use of primitive-based coil systems by providing a generalized approach to their design. We mathematically encode sets of primitive building blocks with spherical harmonic-like symmetries such that all magnetic field variations in free space generated by the sets may be encoded as exact, closed-form expressions. To achieve this, we impose the sets directly into Turner's surface current solution~\cite{10.1088/0022-3727/19/8/001} and apply a spherical harmonic decomposition~\cite{Romeo}. We solve the resulting integral equations analytically to determine the expansion coefficients as simple derivatives with respect to the coil parameters. Then, we use \textit{Mathematica} to construct and simplify the analytic expressions for the expansion coefficients, upon which numerical root-finding routines search the coil parameter space for solutions that cancel field errors. By interpolating these solutions, we determine a meshed contour in the parameter space on which the solutions lie. We then rank the solutions according to a desired attribute, e.g. field fidelity, power-efficiency (field strength per unit current), inductance, or by practical concerns such as spatial extent, overlaps with access holes, or the distance between sets of primitives. We present three worked examples of this process -- one of each using sets of loops, arcs, and ellipses primitives -- to demonstrate the scope of our method, and compare each to standard equivalents. To finish, we design and build uniform axial, linear axial gradient, and quadratic axial gradient, field-generating loops on a 3D-printed cylindrical coil former. We examine how this system can be used to null the background magnetic field in a typical laboratory environment.

The analytic model developed in this work was initially presented at the \textit{Conference on Precision Electromagnetic Measurements 2022}~\cite{CPEMtest}.

\section{Mathematical framework}

\subsection{Field harmonic basis}
In free space, the magnetic field can be represented as the gradient of a magnetic scalar potential, $\mathbf{B}=-\nabla\Psi$. The scalar potential and magnetic field both satisfy Laplace's equation, $\nabla^2\mathbf{B}=\nabla^2\Psi=0$. Here, we express the magnetic scalar potential as the complete basis of real spherical harmonics and so the magnetic field may be expressed in terms of the complete basis of the vector derivatives of each real spherical harmonic~\cite{doi:10.1119/1.1933682}. We refer to this as the basis of \emph{field harmonics},
\begin{equation}\label{eq.bsh}
\mathbf{B}\left(\mathbf{r}\right)=\sum_{n=1}^{\infty} \sum_{m=-n}^{n}\ f_{n,m} \mathbf{B}_{n,m}\left(\mathbf{r}\right),
\end{equation}
where each field harmonic, with magnitude $f_{n,m}$, is denoted by $\mathbf{B}_{n,m}\left(\mathbf{r}\right)=\nabla h_{n,m}\left(\mathbf{r}\right)$ and $h_{n,m}\left(\mathbf{r}\right)$ is the spherical harmonic of the same order $n\in\mathbb{Z}^+$ and degree $m\in\mathbb{Z}\in[-n:n]$. Each spherical harmonic may be expressed as
\begin{align}\label{eq.sph}
     h_{n,m}\left(r,\theta,\phi\right)= \eta_{n,m} r^n P_{n,|m|}\left(\cos\theta\right)
     \begin{pmatrix}
    \cos\left(m\phi\right)\\
    \sin\left(|m|\phi\right)
    \end{pmatrix},
\end{align}
where the upper and lower terms in the right-hand-side bracket denote variations of degree $m\geq0$ and $m<0$, respectively. Each spherical harmonic has a zenith dependence described by Ferrer’s associated Legendre polynomials, $P_{n,|m|}\left(\cos\theta\right)$. The harmonics are defined with the standard normalization,
\begin{align}\label{eq.sphnorm}
\eta_{n,m}&\eta_{n',m'} \int_{0}^{\pi}\mathrm{d}\theta\ \int_{0}^{2\pi}\mathrm{d}\phi\ \ \times \nonumber \\ & \hspace{20pt} h_{n,m}\left(r_0,\theta,\phi\right)h_{n',m'}\left(r_0,\theta,\phi\right) = \delta_{n,n'}\delta_{m,m'},
\end{align}
on a unitary sphere, $r_0=1$, with
\begin{equation}\label{eq.sphnorm2}
\eta_{n,m} = \sqrt{\frac{\zeta_{m,0}\left(2n+1\right)}{4 \pi} \frac{(n-|m|)!}{(n+|m|)!}},
\end{equation}
and $\zeta_{m,m'}=2-\delta_{m,m'}$, where $\delta_{m,m'}$ is the Kronecker delta function.

Here, we shall only consider $\cos\left(m\phi\right)$-like variations, for $m\geq0$. Any $\sin\left(|m|\phi\right)$-like variations, for $m<0$, may be generated by rotating the equivalent $\cos\left(m\phi\right)$-like variation by $\pi/(2|m|)$. The field harmonic components in Cartesian coordinates, $\mathbf{B}_{n,m}\left(\mathbf{r}\right)=X_{n,m}\left(\mathbf{r}\right)\boldsymbol{\hat{x}} + Y_{n,m}\left(\mathbf{r}\right)\boldsymbol{\hat{y}} + Z_{n,m}\left(\mathbf{r}\right)\boldsymbol{\hat{z}}$, are presented in appendix~\ref{app.decomp}. The axial component is
\begin{align}\label{eq.shbz_main}
Z_{n,m}\left(r,\theta,\phi\right)= \eta_{n,m}& \left(n+m\right)r^{n-1} \ \times \nonumber \\
    &\hspace{10pt} P_{n-1,m}\left(\cos\theta\right)
    \cos\left(m\phi\right).
\end{align}
The symmetry of the axial component along the $z$-axis is determined by the parity of $(n+m-1)$ due to its dependence on $P_{n-1,m}\left(\cos\theta\right)$~\cite{mathsbook}. Thus, the axial component is symmetric along the $z$-axis if $(n+m-1)$ is even and is antisymmetric along the $z$-axis if $(n+m-1)$ is odd. 

\begin{figure*}[!ht]
\centering{
\begin{tabular}{c c c}
\includegraphics[page=1]{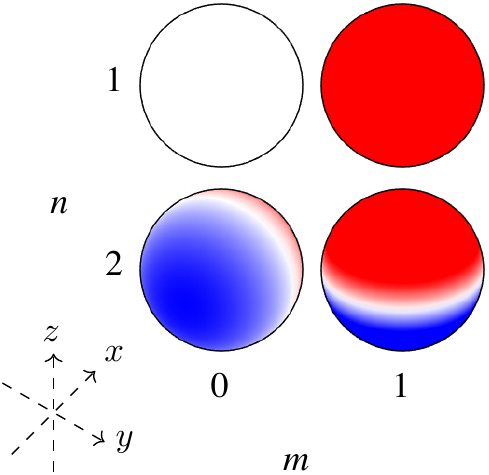}
& \includegraphics[page=2]{pdffigs.pdf} &
\includegraphics[page=3]{pdffigs.pdf} \\
\hspace{27pt}\small{(a)} & \hspace{17pt}\small{(b)} & \hspace{17pt}\small{(c)}
\end{tabular}}
\caption{Cartesian field harmonic components, (a) $X_{n,m}\left(\mathbf{r}\right)$, (b) $Y_{n,m}\left(\mathbf{r}\right)$, and (c) $Z_{n,m}\left(\mathbf{r}\right)$ [equations~\eqref{eq.shbx}--\eqref{eq.shbz}, respectively, in appendix~\ref{app.decomp}] of order $n\in\left[1,2\right]$ and degree $m\in\left[0,1\right]$ shown on the surface of unitary spheres (black outline; red-to-white-to-blue showing positive-to-zero-to-negative field harmonic amplitude), where $\mathbf{B}_{n,m}\left(\mathbf{r}\right)=X_{n,m}\left(\mathbf{r}\right)\boldsymbol{\hat{x}} + Y_{n,m}\left(\mathbf{r}\right)\boldsymbol{\hat{y}} + Z_{n,m}\left(\mathbf{r}\right)\boldsymbol{\hat{z}}$, and $\left(\boldsymbol{\hat{x}},\boldsymbol{\hat{y}},\boldsymbol{\hat{z}}\right)$ represent Cartesian unit vectors.}
\label{fig.fharms}
\end{figure*}

The low order field harmonics encode simple variations when expressed in Cartesian coordinates. For example, for $n=[1,2]$ and $m=[0,1]$ they are
\begin{align}
&\mathbf{B}_{1,0} = \frac{1}{2}\sqrt{\frac{3}{\pi}}\ \boldsymbol{\hat{z}}, \label{eq.B10} \\ 
&\mathbf{B}_{1,1} = \frac{1}{2}\sqrt{\frac{3}{\pi}}\ \boldsymbol{\hat{x}}, \label{eq.B11} \\
&\mathbf{B}_{2,0}\left(x,y,z\right) = \frac{1}{2}\sqrt{\frac{5}{\pi}}\ \left(-x\boldsymbol{\hat{x}}-y\boldsymbol{\hat{y}}+2z\boldsymbol{\hat{z}}\right), \label{eq.B20} \\
&\mathbf{B}_{2,1}\left(x,z\right) = \frac{1}{2}\sqrt{\frac{15}{\pi}}\ \left(z\boldsymbol{\hat{x}}+x\boldsymbol{\hat{z}}\right), \label{eq.B21}
\end{align}
as shown in Fig.~\ref{fig.fharms}.

\subsection{Harmonic matching}
The magnetic field, \eqref{eq.bsh}, generated by the surface currents with $\cos\left(m\phi\right)$-like symmetry flowing on a cylinder of radius $\rho_c$ may be separated into contributions where the axial magnetic field has total symmetry ($+$) or total antisymmetry ($-$) along the $z$-axis about the origin,
\begin{align}
    \mathbf{B}^+\left(\mathbf{r}\right)
    &=\frac{\mu_0}{\pi}\sum^{\infty}_{{\nu}=0}\sum_{m=0}^{\infty}\ \frac{f_{2{\nu}+m+1,m}}{\rho_c^{2{\nu}+m+1}} \mathbf{B}_{2{\nu}+m+1,m}\left(\mathbf{r}\right), \label{eq.Bfinal+} \\
    \mathbf{B}^-\left(\mathbf{r}\right)
    &=\frac{\mu_0}{\pi}\sum^{\infty}_{{\nu}=1}\sum_{m=0}^{\infty}\ \frac{f_{2{\nu}+m,m}}{\rho_c^{2{\nu}+m}} \mathbf{B}_{2{\nu}+m,m}\left(\mathbf{r}\right).\label{eq.Bfinal-}
\end{align}
The axially symmetric field harmonics are of order $n=2{\nu}+m+1$ for $\nu\in\mathbb{Z}^{0+}$ and degree $m\in\mathbb{Z}^{0+}$. The axially antisymmetric field harmonics are of order $n=2{\nu}+m$ for $\nu\in\mathbb{Z}^{+}$ and degree $m\in\mathbb{Z}^{0+}$.

As the magnetic field must relate uniquely to the scalar potential, the spatial forms of the field harmonics are preserved when examining the field generated by the surface current. Thus, the field harmonics must also be present within the well-known Green's function integral expression for the axial field in free space generated inside a cylindrical azimuthal surface current~\cite{10.1088/0022-3727/19/8/001}. We express this in the axially symmetric and antisymmetric cases as
\begin{align}\label{eq.bzdis+}
    B^+_{z}\left(\rho,\phi,z\right)&=-\frac{\mu_0}{\pi\rho_c}\sum_{m=0}^{\infty}\ \zeta_{m,0} \cos\left(m\phi\right) \ \times \nonumber \\ &\hspace{-45pt} \int_{0}^{\infty}\mathrm{d}k \ k\cos\left(\frac{kz}{\rho_c}\right)I_{m}\left(\frac{k\rho}{\rho_c}\right)K'_{m}\left(k\right)J_{\phi}^{m+}\left(\frac{k}{\rho_c}\right),
\end{align}
\begin{align}\label{eq.bzdis-}
    B^-_{z}\left(\rho,\phi,z\right)&=-\frac{i\mu_0}{\pi\rho_c}\sum_{m=0}^{\infty}\ \zeta_{m,0} \cos\left(m\phi\right) \ \times \nonumber \\ &\hspace{-45pt} \int_{0}^{\infty}\mathrm{d}k \ k\sin\left(\frac{kz}{\rho_c}\right)I_{m}\left(\frac{k\rho}{\rho_c}\right)K'_{m}\left(k\right)J_{\phi}^{m-}\left(\frac{k}{\rho_c}\right),
\end{align}
where $I_m\left(z\right)$ and $K_m\left(z\right)$ represent the modified Bessel functions of the first and second kinds, respectively, of order $m$. The Fourier transforms of the axially symmetric and antisymmetric azimuthal current flows, $J_{\phi}^{+}\left(\mathbf{r}'\right)$ and  $J_{\phi}^{-}\left(\mathbf{r}'\right)$, respectively, are
\begin{equation}\label{eq.azft}
    \hspace{-5pt}J_{\phi}^{m\pm}(k) = \frac{1}{2\pi}\int_{0}^{2\pi}\mathrm{d}\phi' \ e^{-im\phi'}\int_{-\infty}^{\infty}\mathrm{d}z'\ e^{-ikz'}J_{\phi}^{\pm}\left(\mathbf{r'}\right).
\end{equation}
As detailed in appendix~\ref{app.match}, we can transform the cylindrical variations in equations~\eqref{eq.bzdis+}~and~\eqref{eq.bzdis-} into spherical variations like those in the axial field harmonic component, \eqref{eq.shbz_main}. We find
\begin{align}\label{eq.cyltosphsym}
     I_{m}\left(\frac{k\rho}{\rho_c}\right) \cos\left(\frac{kz}{\rho_c}\right) &= \sum^{\infty}_{{\nu}=0}\ \frac{(-1)^{{\nu}}}{(2({\nu}+m))!}  \ \times \nonumber \\ &\hspace{10pt} \left(\frac{kr}{\rho_c}\right)^{2{\nu}+m} P_{2{\nu}+m,m}\left(\cos \theta \right),
\end{align}
\begin{align}\label{eq.cyltosphanti}
    I_{m}\left(\frac{k\rho}{\rho_c}\right) \sin\left(\frac{kz}{\rho_c}\right) &= \sum^{\infty}_{{\nu}=1}\ \frac{(-1)^{{\nu}}}{(2({\nu}+m)-1)!} \ \times \nonumber \\ &\hspace{-5pt} \left(\frac{kr}{\rho_c}\right)^{2{\nu}+m-1} P_{2{\nu}+m-1,m}\left(\cos \theta \right).
\end{align}
Substituting equations~\eqref{eq.cyltosphsym}~and~\eqref{eq.cyltosphanti} into equations~\eqref{eq.bzdis+}~and~\eqref{eq.bzdis-} and grouping terms to match the field harmonic basis, \eqref{eq.Bfinal+}~and~\eqref{eq.Bfinal-}, we find
\begin{align}
    f_{2{\nu}+m+1,m} &= \frac{\zeta_{m,0}(-1)^{{\nu}+1}}{(2({\nu}+m)+1)!} \Bigg[ \nonumber \\ & \hspace{10pt} \int_{0}^{\infty}\mathrm{d}k \ k^{2{\nu}+m+1}K'_{m}(k)J_{\phi}^{m+}\left(\frac{k}{\rho_c}\right) \Bigg], \label{eq.fharm+}
\end{align}
\begin{align}
    f_{2{\nu}+m,m} &= \frac{i\zeta_{m,0}(-1)^{{\nu}+1}}{(2({\nu}+m))!} \Bigg[ \nonumber \\ & \hspace{20pt} \int_{0}^{\infty}\mathrm{d}k \ k^{2{\nu}+m}K'_{m}(k)J_{\phi}^{m-}\left(\frac{k}{\rho_c}\right) \Bigg]. \label{eq.fharm-}
\end{align}
The axial field harmonic component is zero for harmonics of equal order and degree, $Z_{n,n}\left(\mathbf{r}\right)=0$. To derive these magnitudes, we follow the same procedure as above using the transverse field, $B_x\left(\mathbf{r}\right)$, from the well-known solution~\cite{10.1088/0022-3727/19/8/001} and the transverse field harmonic component $X_{n,n}\left(\mathbf{r}\right)$. The resulting integrals which determine the field harmonic magnitudes are the same as equations~\eqref{eq.fharm+}~and~\eqref{eq.fharm-}.

\section{Primitive design}
Now, we encode the azimuthal surface current on example sets of simple primitive building blocks: loops, saddles, and ellipses. We design the sets of primitives to have specific azimuthal and axial symmetries so that only certain field harmonics are present in the magnetic field. We then solve the field harmonic magnitude integrals, \eqref{eq.fharm+}~and~\eqref{eq.fharm-}, analytically to exactly determine the magnitudes of the remaining field harmonics.

These solutions are used to choose the coil parameters of combinations of sets of primitives to generate target field harmonics. In example~\ref{example.1}, we design axially antisymmetric loops to generate the $\mathbf{B}_{2,0}$ field harmonic and, in example~\ref{example.23}, we design separate axially symmetric sets of saddles and ellipses to generate the $\mathbf{B}_{2,1}$ field harmonic. The optimization methodology we apply is presented in section~\ref{sec:opt}. The \textit{Mathematica} programs we use to design each coil are publicly-available for non-commercial use and are stored in the repository listed in reference~\cite{NoahRepo}. The default arguments in each of the examples in the repository return the coil systems optimized in the main text.

\subsection{Linear axial gradient with respect to axial position}\label{example.1}
\begin{figure*}[!ht]
\begin{tabular}{c c c}
\includegraphics[page=4]{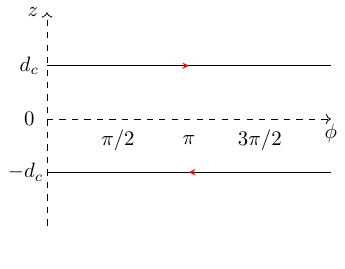}
& \includegraphics[page=5]{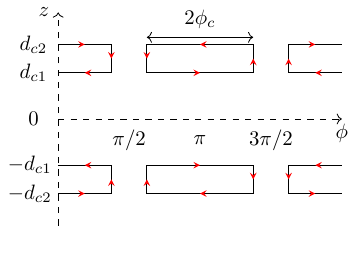} &
\includegraphics[page=6]{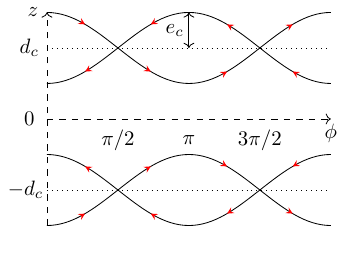} \\
 \hspace{22pt}\small{(a)} & \hspace{22pt}\small{(b)} & \hspace{22pt}\small{(c)}
\end{tabular}
\caption{Variation of primitive current patterns (red arrows indicate current flow direction) in the ${\phi}z$-plane on a cylinder of radius $\rho_c$ to generate the field harmonics optimized in the main text. (a) Axially antisymmetric loops separated axially by $2d_c$ generate field harmonics of order $N=2\nu$, for $\nu\in\mathbb{Z}^{+}$, and degree $M=0$. (b) An even number of pairs of axially symmetric sets of arcs with one-fold symmetry along $\phi$, separated axially by $2d_{c1}$ and $2d_{c2}$ and which extend azimuthally by $2\phi_c$, may be connected to make saddles which generate field harmonics of order $N=2\nu$ for $\nu\in\mathbb{Z}^{+}$ and degree $M=2\mu+1$ for $\mu\in\mathbb{Z}^{+}$. (c) Axially symmetric sets of ellipses with one-fold symmetry along $\phi$, separated axially by $2d_c$ and which extend axially by a maximum of $e_c$, generate the same field harmonics as (b).}
\label{fig.bases}
\end{figure*}
The $\mathbf{B}_{2,0}$ field harmonic is contained within the axially antisymmetric field, \eqref{eq.Bfinal-}, and has $m=0$ azimuthal symmetry (no $\phi$-dependence). As such, it is generated by axially antisymmetric current flows with $m=0$ azimuthal symmetry. These symmetries are matched by pairs of simple loops which carry equal and opposite current, $I$, and are separated about the origin by an axial distance $2d_c$ (Fig.~\ref{fig.bases}a). We may represent such a pair of loops using the following azimuthal current density:
\begin{equation}\label{eq.jphias}
    J^-_{\phi}(z')=I\left(\delta(z'-d_c)-\delta(z'+d_c)\right).
\end{equation}
The Fourier transform, \eqref{eq.azft}, of equation~\eqref{eq.jphias} is
\begin{equation}\label{eq.fppairantim0}
    J_{\phi}^{m-}(k)=-2iI\sin\left(kd_c\right)\delta_{m,0}.
\end{equation}
We determine the field harmonic magnitudes by substituting equation~\eqref{eq.fppairantim0} into equation~\eqref{eq.fharm-},
\begin{align}\label{eq.F_loops_unsolved}
    f_{2{\nu},0}\left(\chi_c\right) &= \frac{2I}{\left(2{\nu}\right)!} \ \times \nonumber \\ & \hspace{20pt} \bigg[(-1)^{{\nu}}  \int_{0}^{\infty} \mathrm{d}k \ k^{2{\nu}} \sin(k\chi_c) K_1(k)\bigg],
\end{align}
where the normalized separation is $\chi_c=d_c/\rho_c$. In appendix~\ref{app.looparcint}, we analytically solve the class of integrals, including equation~\eqref{eq.F_loops_unsolved}, which encode the field harmonic magnitudes generated by loops. We find that
\begin{align}\label{eq.F_loops}
f_{2{\nu},0}\left(\chi_c\right) = \frac{{\pi}I}{\left(2{\nu}\right)!} \frac{\partial^{2{\nu}}}{\partial\chi_c^{2{\nu}}}\left(\frac{\chi_c}{\sqrt{1+\chi_c^2}}\right).
\end{align}

\begin{figure}[!ht]
    \centering
    \begin{tabular}{c c}
    \raisebox{-.5\height}{\scalebox{0.7857}{\includegraphics[page=7]{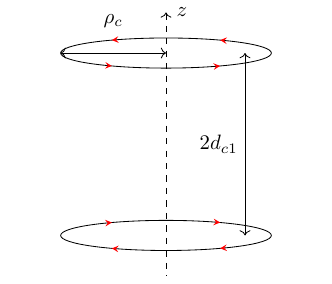}}} & \raisebox{-.5\height}{\scalebox{0.7857}{\includegraphics[page=8]{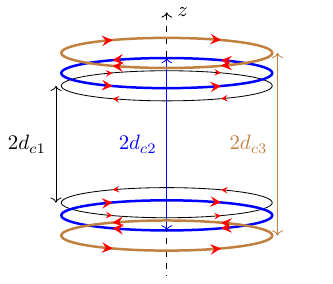}}} \\
    \small{(a)} & \small{(b)}
    \end{tabular}
    \caption{Schematics of $N_{\mathrm{loops}}$ axially antisymmetric loop pairs (red arrows indicate current flow direction) of radius $\rho_c$ at axial positions $z'={\pm}d_{ci}$ for $i\in\left[1:N_{\mathrm{loops}}\right]$. (a) $N_{\mathrm{loops}}=1$ pair of loops with $d_{c1}=\left(\sqrt{3}/2\right)\rho_c$ (black). (b) $N_{\mathrm{loops}}=3$ pairs of loops with $d_{ci} =[0.5544, 0.6748, 0.8660]\rho_c$ and turn ratios $1:-2:2$ (black, blue, and brown, respectively, with higher turn ratios in bold lines).}
    \label{fig.optloops}
\end{figure}
We design $\mathbf{B}_{2,0}$ using 
equation~\eqref{eq.F_loops} by choosing turn ratios and separations to maximize the $f_{2,0}$ field harmonic while minimizing all others. We then proceed by exactly nulling as many \emph{leading-order} field harmonics as possible, which we define as the lowest order field harmonics generated by the selected primitive current pattern that are not the desired field harmonic. This is generally more effective than minimizing as many field harmonics as possible since the leading-order errors have the lowest spatial frequencies. In this specific case, as expected, the leading-order field harmonic, $f_{4,0}$, is nulled exactly for $\chi_c=\sqrt{3}/2$ (Fig.~\ref{fig.optloops}a). This coil geometry may be referred to as an anti-Helmholtz pair/gradient-field Maxwell coil.

Here, we use \textit{Mathematica} (as detailed in section~\ref{sec:opt}) to optimize three axially antisymmetric loop pairs, as presented in Fig.~\ref{fig.optloops}b. We null $\left[f_{4,0},f_{6,0}\right]$ and simultaneously maximize the weighted sum of $f_{2,0}/f_{8,0}$, i.e. the ratio of the target field harmonic magnitude to the magnitude of the next-leading-order-error weighted by the respective turn ratios of each set of primitives. We also constrain all optimized separations to be less than or equal to that of an anti-Helmholtz pair, $\chi_c\leq\sqrt{3}/2$, to make the generated design equally as applicable as an anti-Helmholtz pair in settings where the maximum axial extent of the system is limited, e.g. due to experimental equipment.

\begin{figure}[!ht]
\begin{tabular}{c c}
\includegraphics[page=9]{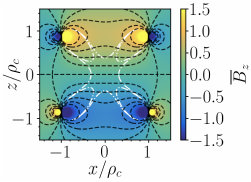} & \includegraphics[page=10]{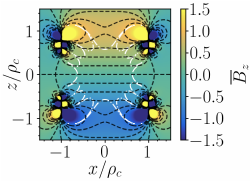} \\
\small{(a)}\hspace{20pt} & \small{(b)}\hspace{20pt}
\end{tabular}
\caption{Magnitude of the normalized axial magnetic field (color scales right), $\overline{B}_z=B_z/B_0$, where $B_0$ is the magnetic field gradient strength specified in Table~\ref{table.stats.discrete}, in the $xz$-plane generated by the coils in Fig.~\ref{fig.optloops}a and Fig.~\ref{fig.optloops}b, corresponding to (a) and (b), respectively, where the coils are of radius $\rho_c$. White contours enclose the regions where $\mathrm{d}B_z/\mathrm{d}z$ deviates from perfect linearity by less than $1$\% (dot-dashed curves).}
\label{fig.B20comp}
\end{figure}
\begin{table*}[!ht]
\caption{Properties of the standard coils in Figs.~\ref{fig.optloops}a,~\ref{fig.optarcs}a~and~\ref{fig.optellipses}a, designed using the methodology presented in Romeo and Hoult~\cite{Romeo}, and the optimized equivalents in Figs.~\ref{fig.optloops}b,~\ref{fig.optarcs}b~and~\ref{fig.optellipses}b designed in this work for a coil radius of $\rho_c=10$~cm. The following coil properties are listed: target field harmonic, $\mathbf{B}_{n,m}$ of order $n$ and degree $m$; the maximum axial dimension, $\mathrm{max}\left(\mathrm{dim}_c\right)$, where $\mathrm{dim}_c=d_c$ in the loops and saddles cases, $\mathrm{dim}_c=d_c+e_c$ in the ellipses case, $d_c$ is the separation of the coil units from the origin, and $e_c$ is the axial extent of the ellipses from their centre; the total length of wire in the coil, $l$, including repeated units with multiple turn ratios; the magnetic field gradient strength, $B_0$, i.e. $\mathrm{d}B_z/\mathrm{d}z$ for $\mathbf{B}_{2,0}$ and $\mathrm{d}B_x/\mathrm{d}z$ for $\mathbf{B}_{2,1}$ along the $z$-axis at the centre of the coil, per unit current, $I$; the inductance, $L$; and the size of the central volume where the target field gradient is generated with less than $1$\% deviation from target relative to that generated by the standard coil, $V_{1\%}$. The inductance is approximated numerically from the total magnetic energy~\cite{PurcellEM} using COMSOL Multiphysics\textsuperscript{\textregistered}, assuming that the wire tracks are filamentary.}
\label{table.stats.discrete}
    \centering
    {\setlength{\arrayrulewidth}{0.25mm}
    \setlength{\tabcolsep}{2pt}
    \renewcommand{\arraystretch}{1.1}
    \begin{tabular}{|P{0.09\columnwidth} | P{0.3\columnwidth} || P{0.3\columnwidth} | P{0.3\columnwidth} | P{0.3\columnwidth} | P{0.3\columnwidth} | P{0.3\columnwidth}|}
    \hline
    $\mathbf{B}_{n,m}$ & Coil &  $\mathrm{max}\left(\mathrm{dim}_c\right)$, (cm) & $l$, (m) & $B_0/{I}$, ($\mu$T/Am) & $L$, ($\mu$H) & $V_{1\%}$ \\
    \hline \hline
    \multirow{2}{*}{$\mathbf{B}_{2,0}$} & Standard, Fig.~\ref{fig.optloops}a & $8.66$ & $1.26$ & $80.6$ & $1.07$ & $1$ \\ 
    & Optimized, Fig.~\ref{fig.optloops}b & $8.66$ & $6.28$ & $69.0$ & $5.61$ & $2.85$ \\
    \hline \hline
    \multirow{4}{*}{$\mathbf{B}_{2,1}$} & Standard, Fig.~\ref{fig.optarcs}a & $25.6$ & $3.40$ & $89.2$ & $2.59$ & $1$  \\
    & Optimized, Fig.~\ref{fig.optarcs}b & $25.5$ & $9.93$ & $131$ & $11.0$ & $4.55$  \\
    & Standard, Fig.~\ref{fig.optellipses}a & $22.5$ & $3.06$  & $95.0$ & $2.85$ & $1$  \\
    & Optimized, Fig.~\ref{fig.optellipses}b & $11.5$ & $5.18$ & $16.8$ & $1.40$ & $6.26$  \\
    \hline
\end{tabular}}
\end{table*}
In Fig.~\ref{fig.B20comp}, the axial magnetic field generated by an anti-Helmholtz pair is compared to that generated by the optimized coil. The properties of the coils are summarized in Table~\ref{table.stats.discrete}. The size of the central volume where $\mathbf{B}_{2,0}$ is generated with less than $1$\% error (less than $1$\% deviation from target; bounded within the central dot-dashed curves in Fig.~\ref{fig.B20comp}) is a factor of $2.85$ greater than that generated by an anti-Helmholtz pair. However, due to the increased number of turns and alternating turn ratio polarities, compared to the anti-Helmholtz pair, the optimized coil is $1.17$ times less power-efficient and has $5.24$ times greater inductance. In a scenario where fast current switching is required alongside high gradient linearity, the inductance could be calculated analytically~\cite{Turner_1988} and imposed as an additional minimization condition.

\subsection{Linear transverse gradient with respect to axial position}\label{example.23}
The $\mathbf{B}_{2,1}$ field harmonic is contained within the axially symmetric field, \eqref{eq.Bfinal+}, and has $m=1$ azimuthal symmetry (one line of symmetry in the $\rho\phi$-plane). To match this, we optimize sets of symmetric arcs of azimuthal extent $2\phi_c$, which are axially separated about the origin by $2d_c$ and have one-fold azimuthal periodicity (the arc pairs repeat every $\phi=\pi$ with alternating polarity; Fig.~\ref{fig.bases}b). The azimuthal current density that describes one such set of arcs may be represented as 
\begin{align}\label{eq.jphiarcs}
    J^+_{\phi}(\phi',z') &= I\left(\delta(z'-d_c)+\delta(z'+d_c)\right) \ \times \nonumber \\ & \hspace{-40pt} \sum_{\lambda=0}^{1}\ (-1)^{\lambda}\left[ H\left(\phi'+\phi_c-\lambda\pi\right)-H\left(\phi'-\phi_c-\lambda\pi\right)\right],
\end{align}
where $H\left(x\right)$ is the Heaviside step function. The Fourier transform, \eqref{eq.azft}, of equation~\eqref{eq.jphiarcs} is
\begin{align}\label{eq.fppairsym}
    J_{\phi}^{m+}(k)&=\frac{2I(1-(-1)^m)}{\pi m}\sin(m\phi_c)\cos\left(kd_c\right).
\end{align}
Substituting equation~\eqref{eq.fppairsym} into equation~\eqref{eq.fharm+}, we find
\begin{align}\label{eq.F_arcs_pre}
    f_{2\nu+m+1,m}\left(\chi_c,\phi_c\right) &= \frac{4I(1-(-1)^m)}{{\pi}m\left(2\left(\nu+m\right)+1\right)!}\sin(m\phi_c) \ \times \hspace{120pt} \nonumber \\ & \hspace{-60pt} \bigg[(-1)^{{\nu+1}}\int_{0}^{\infty} \mathrm{d}k \ k^{2\nu+m+1} \cos(k\chi_c) K'_{m}(k)\bigg].
\end{align}
As with the loops case, we analytically solve the class of integrals, including equation~\eqref{eq.F_arcs_pre}, which encode the field harmonic magnitudes generated by arcs in appendix~\ref{app.looparcint}. We find
\begin{align}\label{eq.F_arcs}
f_{2\nu+m+1,m}\left(\chi_c,\phi_c\right) &= \frac{I(2m)!(1-(-1)^m)}{2^{m-1}m!\left(2\left(\nu+m\right)+1\right)!} \ \times \nonumber \\ & \hspace{-50pt} \sin(m\phi_c) \frac{\partial^{2\nu+1}}{\partial\chi_c^{2\nu+1}} \Bigg[ \frac{\chi_c}{m}\left(\frac{1}{1+\chi_c^2}\right)^{m+1/2} \ + \nonumber \\ & \hspace{-40pt} \sum_{k=0}^{m-1} \frac{(-1)^{k}}{2k+1} {\binom{m-1}{k}} \left(\frac{\chi_c^2}{1+\chi_c^2}\right)^{k+1/2} \Bigg],
\end{align}
where the binomial coefficient is $\binom{n}{k}=n!/\left(\left(n-k\right)!k!\right)$.

Now, we use equation~\eqref{eq.F_arcs} to select turn ratios, separations, and extents of four sets of arcs to maximize $f_{2,1}$ while minimizing leading-order errors. To obey current continuity, the number of sets of arcs must be even and the current in each set of arcs must correspond to an equal and opposite current in another set of arcs. As a result, we join each set of arcs to another to form double saddles (Fig.~\ref{fig.bases}b).

\begin{figure}[!ht]
    \centering
    \begin{tabular}{c c}
    \raisebox{-.5\height}{\scalebox{0.7857}{\includegraphics[page=11]{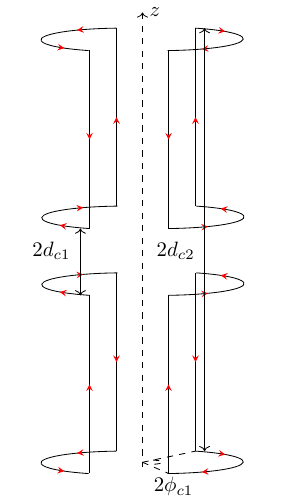}}} & \raisebox{-.5\height}{\scalebox{0.7857}{\includegraphics[page=12]{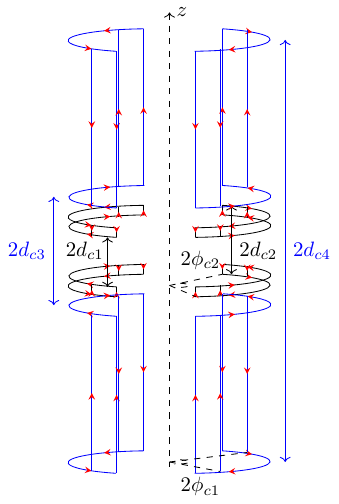}}} \\
    \small{(a)} & \small{(b)}
    \end{tabular}
    \caption{Schematics of $N_{\mathrm{arcs}}$ axially symmetric sets of arcs (red arrows indicate current flow direction) with one-fold azimuthal symmetry of radius $\rho_c$ at axial positions $z'={\pm}d_{ci}$ for $i\in\left[1:N_{\mathrm{arcs}}\right]$ which are connected to make double saddles. (a) $N_{\mathrm{arcs}}=2$ sets of arcs (black) which extend azimuthally over $2\phi_{c1}=2\pi/3$ and are at axial positions $d_{ci} =[0.404, 2.56]\rho_c$. The sets of arcs are connected in series to make double saddles. (b) $N_{\mathrm{arcs}}=4$ sets of arcs, each of which contains two nested azimuthal extensions, $2\phi_{c1}=7\pi/15$ and $2\phi_{c2}=13\pi/15$, at axial positions $d_{ci}=[0.2995, 0.4170, \allowbreak 0.6550, 2.5522]\rho_c$, where the first and second (black) and the third and fourth (blue) sets of arcs are connected in series to make double saddles with the same turn ratios.}
    \label{fig.optarcs}
\end{figure}%
Firstly, we note that equation~\eqref{eq.F_arcs} prohibits all even degrees and shows us that all field harmonics of odd degrees are proportional to $\sin\left(m\phi_c\right)$. We find optimal extents of multiple sets of arcs by substituting $\sin\left(m\phi_c\right)$ for the relevant Chebyshev polynomial~\cite{mathsbook}, and then finding the root of the functions where leading-order degrees are nulled. Here, to null all field harmonics with $m=\left[3,5\right]$, the required azimuthal extents of two sets of arcs with unitary turn ratios are $\phi_c=[7\pi/30,13\pi/30]$. We then use \textit{Mathematica} to optimize four sets of arcs (each of which has two nested azimuthal extensions) to null the leading-order errors, $\left[f_{4,1},f_{6,1},f_{8,1}\right]$, and maximize the weighted sum of $f_{2,1}/f_{10,1}$ generated by the four sets of arcs. As with example~\ref{example.1}, we also constrain all axial separations, here to $\chi_c\leq2.56$, so that the maximum extent is less than or equal to that of a standard design created in reference~\cite{Romeo} (Fig.~\ref{fig.optarcs}a).

\begin{figure}[!ht]
\begin{tabular}{c c}
\includegraphics[page=13]{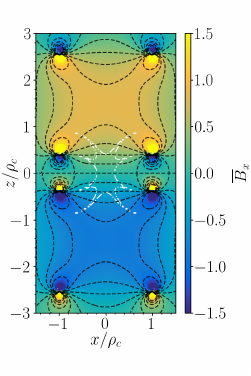} & \includegraphics[page=14]{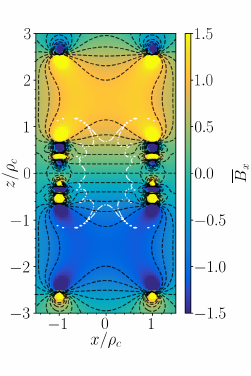} \\
\small{(a)}\hspace{20pt} & \small{(b)}\hspace{20pt}
\end{tabular}
\caption{Magnitude of the normalized transverse magnetic field (color scales right), $\overline{B}_x=B_x/B_0$, where $B_0$ is the magnetic field gradient strength specified in Table~\ref{table.stats.discrete},  in the $xz$-plane generated by the coils in Fig.~\ref{fig.optarcs}a and Fig.~\ref{fig.optarcs}b, corresponding to (a) and (b), respectively, where the coils are of radius $\rho_c$. White contours enclose the regions where $\mathrm{d}B_x/\mathrm{d}z$ deviates from perfect linearity by less than $1$\% (dot-dashed curves).}
\label{fig.B21comparcs}
\end{figure}
The optimized coil design is presented in Fig.~\ref{fig.optarcs}b. In Fig.~\ref{fig.B21comparcs} we show that the optimized coil generates $\mathbf{B}_{2,1}$ much more effectively than the standard coil. The power-efficiency of the optimized coil is $1.47$ times greater and it generates $\mathbf{B}_{2,1}$ with less than $1$\% error over a central region that is $4.55$ times greater in volume (Table~\ref{table.stats.discrete}). Compared with the standard coil, however, as with example~\ref{example.1}, the increased complexity of the optimized coil arrangement increases the comparative inductance by a factor of $4.25$.

Given the acute changes in flow direction in saddles between current-carrying arcs and axial connecting wires, inaccuracy in saddle construction may generate unwanted high-order field harmonics~\cite{Romeo}. To overcome this, we can use smoothly-varying elliptical wire tracks on the surface of a cylinder instead of saddles to generate the same field harmonics. 

Here, we demonstrate this by generating $\mathbf{B}_{2,1}$ using sets of axially symmetric cylindrical ellipses, which each extend over an axial distance $2e_c$ and are separated about the origin by a central axial distance $2d_c>2e_c$. The sets of ellipses do not cross, preserving the axial symmetry, and have one-fold azimuthal periodicity (the ellipse pairs repeat every $\phi=\pi$ with alternating polarity; Fig.~\ref{fig.bases}c). The azimuthal current density which traces the path of one such set of ellipses may be represented as
\begin{align}\label{eq.jphiell}
    J^+_{\phi}(\phi',z') &=
     I \sum_{\lambda=0}^{1}\ (-1)^{\lambda}\bigg[  \nonumber \\ & \hspace{20pt} \delta\left(z'-\left(d_c+e_c\cos\left(\phi'-\lambda\pi\right)\right)\right)+ \nonumber  \\ & \hspace{40pt} \delta\left(z'+\left(d_c+e_c\cos\left(\phi'-\lambda\pi\right)\right)\right)\bigg].
\end{align}
The Fourier transform, \eqref{eq.azft}, of equation~\eqref{eq.jphiell} is
\begin{equation}\label{eq.fpellsym}
    J_{\phi}^{m+}(k)=-2Ii^{m+1}(1-(-1)^m) J_{m}\left(ke_c\right)\sin\left(kd_c\right),
\end{equation}
where $J_{m}\left(z\right)$ represents the Bessel function of the first kind of order $m$. Substituting equation~\eqref{eq.fpellsym} into equation~\eqref{eq.fharm+}, we find
\begin{align}\label{eq.F_ellipses_pre}
    f_{2\nu+m+1,m}\left(\chi_c,\psi_c\right) &= \frac{4I(1-(-1)^m)}{\left(2\left(\nu+m\right)+1\right)!} \bigg[(-1)^{\nu+(m+3)/2} \ \times \hspace{120pt} \nonumber \\ & \hspace{-50pt} \int_{0}^{\infty} \mathrm{d}k \ k^{2\nu+m+1} J_{m}(k\psi_c) \sin(k\chi_c) K'_{m}(k)\bigg],
\end{align}
where the normalized extent is $\psi_c=e_c/\rho_c$.

We solve the class of integrals which include equation~\eqref{eq.F_ellipses_pre} analytically in appendix~\ref{app.ellipseint}. We find
\begin{align}\label{eq.F_ellipses}
    f_{2\nu+m+1,m}\left(\chi_c,e_c\right) &= \frac{8I}{\left(2\left(\nu+m\right)+1\right)!} (-1)^{m+1} \ \times \nonumber \\ &\hspace{-40pt} \frac{\partial^{2\nu+1}}{\partial\chi_c^{2\nu+1}}\Bigg[  \chi_c \frac{\partial^{m} \tilde{S}_{m}\left(\chi_c,\psi_c\right)}{\partial\chi_c^{m}} \ + \hspace{200pt} \nonumber \\ &\hspace{-40pt} m \frac{\partial^{m-1} \tilde{S}_{m}\left(\chi_c,\psi_c\right)}{\partial\chi_c^{m-1}} + \psi_c \frac{\partial^{m} \tilde{S}_{m}\left(\chi_c,\psi_c\right)}{\partial\chi_c^{m-1}\partial\psi_c}\Bigg],
\end{align}
where
\begin{align}\label{eq.S_ellipses}
\tilde{S}_m\left(\chi_c,\psi_c\right) = \frac{(1-i)}{2i^m\sqrt{2\psi_c}} Q_{m-1/2}\left(i\frac{\psi_c^2-\chi_c^2-1}{2\psi_c}\right),
\end{align}
and $Q_{m-1/2}\left(z\right)$ represents a Legendre function of the second kind of half-integer order, $\left(m-1/2\right)$.

Now, we use equation~\eqref{eq.F_ellipses} to select turn ratios, separations, and extents of two sets of ellipses to maximize $f_{2,1}$ while minimizing leading-order errors. Unlike the saddles case, the optimization of orders and degrees cannot be separated. Thus, to maximize a field harmonic using two sets of ellipses, we completely null the leading-order magnitudes, $\left[f_{4,1},f_{4,3},f_{6,1}\right]$, and maximize the weighted ratio of $f_{2,1}/f_{8,1}$ generated by both sets of ellipses. Additionally, we impose $\left(\chi_c+\psi_c\right)\leq1.32$ to limit the maximum system dimension to be less than or equal to half that of the optimized saddle coils.

\begin{figure}[!ht]
    \centering
    \begin{tabular}{c c}
    \raisebox{-.5\height}{\scalebox{0.7857}{\includegraphics[page=15]{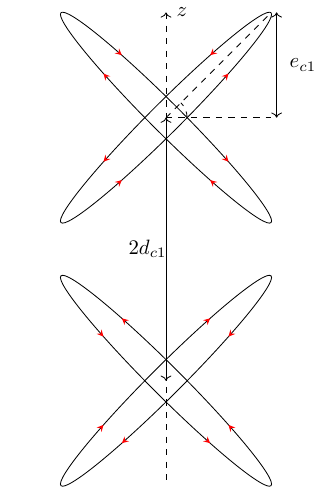}}} & \raisebox{-.5\height}{\scalebox{0.7857}{\includegraphics[page=16]{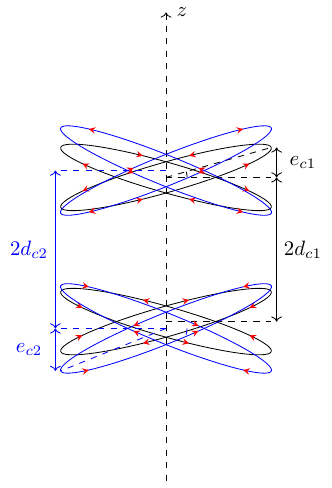}}} \\
    \small{(a)} & \small{(b)}
    \end{tabular}
    \caption{Schematics of $N_{\mathrm{ellipses}}$ axially symmetric sets of ellipses (red arrows indicate current flow direction) with one-fold azimuthal symmetry of radius $\rho_c$ at axial positions $z'={\pm}d_{ci}$ for $i\in\left[1:N_{\mathrm{ellipses}}\right]$. (a) $N_{\mathrm{ellipses}}=1$ set of ellipses with $d_{c1} =1.25\rho_c$ and which extend by a maximum axial distance $e_{c1}=\rho_c$ (black). (b) $N_{\mathrm{ellipses}}=2$ sets of ellipses with $d_{ci} =[0.6842, 0.7493]\rho_c$ which extend by maximum axial distances $e_{ci}=[0.2842, 0.4036]\rho_c$ and have opposite current ratios, $1:-1$ (black and blue, respectively).}
    \label{fig.optellipses}
\end{figure}%
\begin{figure}[!ht]
\begin{tabular}{c c}
\includegraphics[page=17]{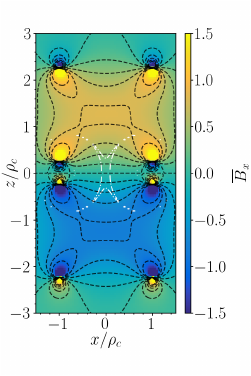} & \includegraphics[page=18]{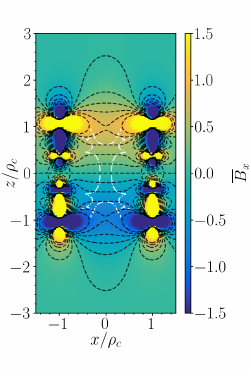} \\
\small{(a)}\hspace{20pt} & \small{(b)}\hspace{20pt}
\end{tabular}
\caption{Magnitude of the normalized transverse magnetic field (color scales right), $\overline{B}_x=B_x/B_0$, where $B_0$ is the magnetic field gradient strength specified in Table~\ref{table.stats.discrete},  in the $xz$-plane generated by the coils in Fig.~\ref{fig.optellipses}a and Fig.~\ref{fig.optellipses}b, corresponding to (a) and (b), respectively, where the coils are of radius $\rho_c$. White contours enclose the regions where $\mathrm{d}B_x/\mathrm{d}z$ deviates from perfect linearity by less than $1$\% (dot-dashed curves).}
\label{fig.B21compellipses}
\end{figure}
In Figs.~\ref{fig.optellipses}~and~\ref{fig.B21compellipses} we present schematics and performances of a standard coil design from reference~\cite{Romeo} and the optimized design, respectively. As expected, the optimized elliptical coils are effective at generating $\mathbf{B}_{2,1}$ even with the constraint on system extent. Compared to the standard double saddles (Fig.~\ref{fig.optarcs}a) and ellipses (Fig.~\ref{fig.optellipses}a), the optimized elliptical coil has a maximum axial extent $2.22$ and $1.96$ times smaller, but generates $\mathbf{B}_{2,1}$ with less than $1$\% error over central volumes $1.36$ and $6.26$ times greater in size. However, due to opposite directions of current polarity between the two sets of ellipses in the optimized design, the standard coils are $5.31$ and $5.65$ times more power-efficient than the optimized elliptical coil.

\section{Optimization routine}\label{sec:opt}
\begin{figure}[!ht]
    \centering
    \includegraphics[page=19]{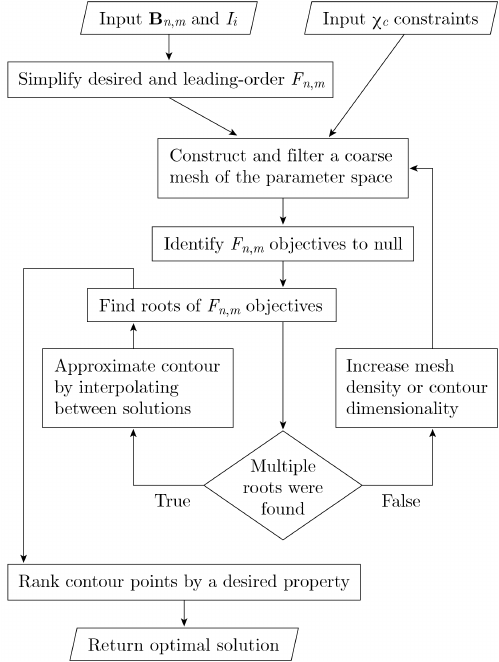}
    \caption{The algorithm used to determine optimal sets of normalized coil separations, $\chi_{c}$. The inputs are: a target field harmonic, $\mathbf{B}_{n,m}$, of order $n$ and degree $m$; turn ratios of primitive groups, $I_{i}$; and constraints on the values of $\chi_{c}$. The weighted magnitudes of field harmonics, $F_{n,m}$, are used to search the parameter space for solutions on the contours where the set of leading-order harmonics are nulled. These solutions are ranked and the optimal solution is returned.}
    \label{fig.algorithm}
\end{figure}

Our goal is to find an \emph{optimal} set of coil parameters which maximize a desired field harmonic while minimizing unwanted contributions. We shall only consider optimization of continuous geometric parameters, not discrete turn ratios which we specify alongside other properties like the minimum separation between primitives. To determine the best turn ratios, we re-run the optimization for simple combinations of turn ratios and then rank solutions across all runs.

Each coil parameter affords a degree of freedom with which to null a field harmonic. In theory, it is possible to null as many field harmonics as there are coil parameters, but in non-trivial scenarios these solutions are difficult to find and may not exist within the constrained parameter space. Instead, we consider an underdetermined problem system, i.e. we null fewer field harmonics than there are coil parameters. In the space of coil parameters, the solutions lie on a contour of dimension equal to the number of coil parameters less the number of nulled harmonics. Finding a solution on this contour using numerical techniques is faster and more reliable than finding a single-point solution. Increasing the dimensionality of the contour makes it easier to find solutions at the expense of nulling fewer harmonics. Once solutions are found on a contour, we rank them according to a desired property, such as the ratio of the desired-to-leading-order-error harmonic magnitudes, and then select the best-ranked solution. We implement the optimization in \textit{Mathematica} because of its fast and easy-to-use symbolic expression simplification (via the \texttt{Simplify} function), numerical root-finding (via the \texttt{FindRoot} function), and process parallelization.

Let us consider the design presented in subsection~\ref{example.1} to generate $\mathbf{B}_{2,0}$ using $N_\mathrm{loops}=3$ pairs of loops. Our optimization algorithm is summarized in Fig.~\ref{fig.algorithm}. 

\begin{figure}[htb!]
\centering
\begin{tabular}{c c}
\includegraphics[page=20]{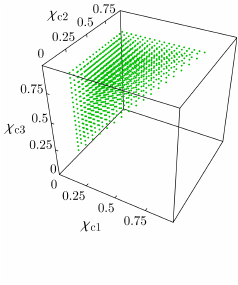} & \includegraphics[page=21]{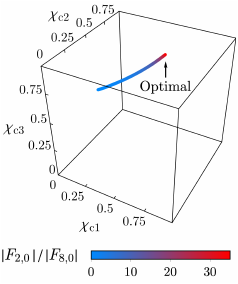} \\
\hspace{5pt}\small{(a)} & \hspace{5pt}\small{(b)}
\end{tabular}
\caption{(a) Initial seeds (green scatter) for the \texttt{FindRoot} function in \textit{Mathematica} in the search space of optimal normalized separations, $\chi_{ci}$, for $i{\in}[1:N_\mathrm{loops}]$, to design the set of $N_\mathrm{loops}=3$ axially antisymmetric loop pairs presented Fig.~\ref{fig.optloops}b. (b) Contour of solutions in the search space of $\chi_{ci}$ that null the field harmonics of undesired orders $n=\left[4,6\right]$ and degree $m=0$. Coloring corresponds to the magnitude of the ratio of the weighted sum of the magnitudes of the target field harmonic generated by the set of loop pairs, $F_{2,0}$, to that of the leading-order error field harmonic generated by the set of loop pairs, $F_{8,0}$. The optimal solution (arrow) maximizes this ratio.}
\label{fig.discretemethod}
\end{figure}
Firstly, we construct and simplify symbolic expressions for the total magnitudes of the low-order field harmonics generated by the primitive set, weighted by the predetermined turn ratios. The weighted magnitude of each field harmonic is
\begin{equation} \label{eq.Fweight}
F_{n,0}=\sum_{i=1}^{N_\mathrm{loops}}\ I_{i} f_{n,0}\left(\chi_{ci}\right),
\end{equation}
where $I_i$ are the turn ratios of each loops pair and the individual field harmonic magnitudes, $f_{n,0}$, may be obtained using equation~\eqref{eq.F_loops}. We use equation~\eqref{eq.Fweight} as the objective function to search for solutions of $F_{4,0}=F_{6,0}=0$ in a coarse mesh of the $N_\mathrm{loops}$\nobreakdash-dimensional parameter space. In this design, three filtering conditions are applied to the mesh to bound the search space and impose a minimum distance between adjacent coil pairs: $\chi_{ci}\geq0.1$, $\chi_{ci}\leq\sqrt{3}/2$, and $\left(\chi_{c(i+1)}-\chi_{ci}\right)\geq0.01$. This reduces the number of mesh points to $1140$ (Fig.~\ref{fig.discretemethod}a). 

At each mesh point, we apply \texttt{FindRoot}, which uses a Newton--Raphson method with step control~\cite{doi:10.1137/1.9781611971200.ch6} to search locally around positions in the solution space for locations where the undesired harmonics are nulled. The search over $1140$ points takes $0.65$ seconds to complete\footnote{Calculations are bench-marked using a \textit{MacBookPro18,2} equipped with a \textit{M1~Max} processor with eight 3228~MHz \textit{performance} cores, two 2064~MHz \textit{efficiency} cores, and 32~GB of 512-bit LPDDR5 SDRAM memory.} and $21$ unique points are found on the one-dimensional solution contour. We then linearly interpolate between the known points on the solution contour to estimate further solutions. The interpolated points are randomized slightly to expand the evaluation scope, and are then used to seed \texttt{FindRoot} once more. This process occurs at $411$ interpolated points and takes $0.18$ seconds. This is more efficient than the previous step because the seeds are close to the solution contour and therefore converge rapidly. In this example, all the interpolated seeds converge onto the solution contour, which is shown in Fig.~\ref{fig.discretemethod}b, but in cases where the contour is discontinuous or varies greatly, not all seeds may converge. Finally, we rank the solutions on the contour according to the absolute value of the ratio of the magnitude of the desired field harmonic, $F_{2,0}$, to the leading-order error field harmonic, $F_{8,0}$. We select the solution that maximizes this ratio (denoted with an arrow in Fig.~\ref{fig.discretemethod}b), which takes $0.01$ seconds for the $\mathbf{B}_{2,0}$ example.

\section{Case study: Axial nulling}\label{sec:case}
Now, we present a case study demonstrating the implementation of optimized magnetic field coils to null residual axial variations in free space, e.g. for residual field compensation for atomic magnetometers~\cite{9743468} or atom interferometers~\cite{Hobson_2022}.

\begin{figure}[!ht]
    \centering
    \includegraphics[page=22]{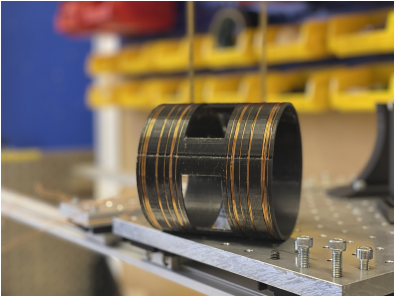}
    \caption{Hand-wound 3D-printed coil former containing independent uniform axial, $B_z$, linear axial gradient, $\mathrm{d}B_z/\mathrm{d}z$, and quadratic axial gradient, $\mathrm{d}^2B_z/\mathrm{d}z^2$, field-generating coils.}
    \label{fig.coilformer}
\end{figure}%
The system comprises the loop coil in Fig.~\ref{fig.optloops}b which generates the $\mathbf{B}_{2,0}$ field harmonic, along with other loop coils (presented in  appendix~\ref{app.designs}) designed using our open-access \textit{Mathematica} program to generate the $\mathbf{B}_{1,0}$ and $\mathbf{B}_{3,0}$ field harmonics. We choose to generate these field harmonics as they are the lowest order $m=0$ field harmonics~\cite{Romeo}, and so are most likely to be present in the background field. The coils are wound by hand using $0.63$~mm diameter ($22$~AWG) enamelled copper wire on a coil former of radius of $\rho_c=50$~mm (Fig.~\ref{fig.coilformer}). During optimization, the normalized axial separations are bounded between $0.35<\chi_c<1$. This allows the coil former to be only $100$~mm long and leaves a large central axial region of length $35$~mm where there are no wires. This area contains four optical access holes that are evenly spaced and have a height of $30$~mm and an azimuthal extent of $\pi/3$ radians. The coil former is manufactured using Fused Deposition Modeling (FDM) with a commercially available UltiMaker S5 using hard PLA body material and PVA support material. The PVA support is dissolved during post-processing by soaking the print in lukewarm water for $48$ hours. The FDM printer is capable of rendering grooves in the coil former with a resolution of $0.06$~mm and uses cost-effective materials ($<\textnormal{£}20$ in this scenario).

\begin{figure}[!ht]
\centering
\includegraphics[page=23]{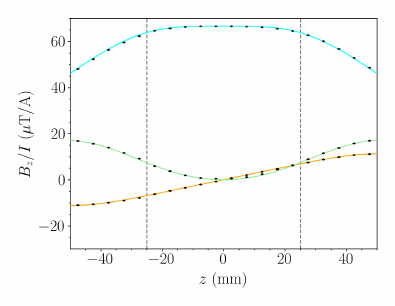} \\    
\hspace{5pt}\small{(a)} \\
\includegraphics[page=24]{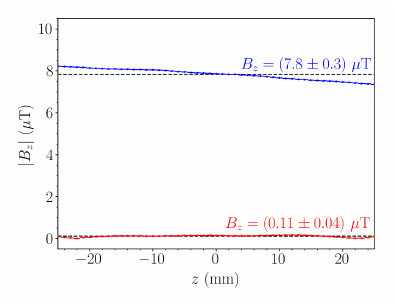} \\    
\hspace{5pt}\small{(b)}
\caption{(a) Magnitude of the axial magnetic field, $B_z$, per unit current, $I$, along the $z$-axis, generated by the uniform (light blue curve), linear gradient (orange curve), and quadratic gradient (light green curve) axial field-generating coils presented in Fig.~\ref{fig.optloopsbias}a, Fig.~\ref{fig.optloops}b, and Fig.~\ref{fig.optloopsbias}b, alongside experimental measurements (black scatter). The magnetic null region is shown with dashed lines (grey); (b) Measured static axial magnetic field magnitude pre- (blue curve) and post- (red curve) active nulling, alongside labels (blue and red) displaying the mean (also shown with dashed black curves) and standard deviation in the axial field magnitude within the null region.}
\label{fig.biasmeasurenull}
\end{figure}
The axial magnetic field generated by each coil is measured using a Stefan Mayer Fluxmaster magnetometer connected to a NI-USB 6212 data acquisition system.  During each measurement, sinusoidal currents of amplitude $80$~mA and oscillating at a frequency of $5$~Hz are passed through each coil in turn for $7$~s. The magnetic field produced by each coil is obtained by calculating the Fast Fourier Transform (FFT) of the magnetometer output, eliminating the need for separate measurement offsets. The measured and expected field profiles show excellent agreement and are presented in Fig.~\ref{fig.biasmeasurenull}a.

Next, the static axial magnetic field along the central $50$~mm of the coil's axis is decomposed into uniform, linear gradient, and quadratic gradient contributions using the method of least-squares, following the method in Ref.~\cite{10.48550/arxiv.2210.15612}. Coil currents of $I=\left(\left[178,59,9\right]{\pm}1\right)$~mA are applied to the $\mathbf{B}_{1,0}$, $\mathbf{B}_{2,0}$, and $\mathbf{B}_{3,0}$ field-generating coils, respectively, to null the measured static background. The magnitude of the axial field pre- and post- null is shown in Fig.~\ref{fig.biasmeasurenull}b. Over the central $50$~mm of the coil's axis, the mean axial field magnitude reduces by over a factor of $70$ from $7.8$~$\mu$T to $0.11$~$\mu$T and the standard deviation in the axial field magnitude reduces by over a factor of $7$ from $300$~nT to $40$~nT. To further diminish the axial field, several approaches could be considered, such as adding more field-generating coils to null higher order field harmonics, increasing the number of leading-orders nulled in existing coil designs by increasing the number of separations optimized, or incorporating dynamic feedback from a reference magnetometer to update the applied coil currents~\cite{10.1063/1.5087957}.

\section{Conclusion and outlook}
Here, we have presented a spherical harmonic decomposition of the magnetic field in free space generated inside sets of primitive structures based on loops, saddles, and ellipses. In each case, we solved for the magnetic field harmonic magnitudes as simple derivatives with respect to the coil parameters. These derivatives can be computed rapidly using computer algebra software, and in this work we used \textit{Mathematica}. We then demonstrated a generalized approach to design simple primitive coil structures by mapping the landscape of parameters which control the geometry of each set. Using this approach, we designed three field-generating structures -- nested sets of loops, arcs, and ellipses -- which generate target fields accurately by completely nulling a set number of leading-order undesired contributions. To allow fair comparison with standard systems, in each worked example we chose the solutions closest to the global optimum (which null the leading-order deviations while maximizing the target-field-to-next-leading-order-error ratio) within the constrained space of coil parameters. However, the optimization procedure could be readily adjusted to prioritize other constraints such as power-efficiency, size, and inductance. Furthermore, more diverse objective functions could also be added to ensure that designs can be manufactured easily, e.g. sensitivity to wire misplacement or number of wire overlaps. 

Although we have focused on specific worked examples within the main text, the \textit{Mathematica} program we provide~\cite{NoahRepo} may be used to design any low-order field harmonic using the integral solutions in appendices~\ref{app.looparcint}~and~\ref{app.ellipseint}. We use the loops coil optimized in the main text, alongside other loops coils generated using our program (see appendix~\ref{app.designs}), to construct a 3D-printed, hand-wound axial field compensation system. By calculating the appropriate currents applied to the coil, we were able to significantly reduce both the magnitude and variation in the measured axial field along the central half of the coil's axis in a typical laboratory environment. Specifically, we were able to reduce the axial field magnitude from $7.8$~$\mu$T to $0.11$~$\mu$T and its standard deviation from $300$~nT to $40$~nT. 

One can design any magnetic field in free space using our program by decomposing the target field into a weighted sum of field harmonics and combining sets of axially symmetric and antisymmetric primitives. Optimized arrangements of primitive coils may be useful in diverse contexts such as the design of accurate and power-efficient linear gradient fields for magneto--optical traps~\cite{doi:10.1063/5.0011428} or the cancellation of external interference for atomic magnetometers~\cite{10.48550/arxiv.2210.15612} by using modified governing equations accounting for the response of external passive magnetic shielding~\cite{DiscretePaper}. The field harmonics generated by primitives could also be posed and solved for other surface geometries. For example, cuboidal primitives may be useful for generating magnetically-shielded enclosures~\cite{10.1038/s41598-022-17346-1} and toroidal primitives may be applied for plasma confinement in tokamaks~\cite{VOSS2000407}. Discrete building block and discretized surface current-based coils could also be implemented together to maximize their performance, e.g. broadband reduction in the noise floor using power-efficient simple building block coils combined with accurate shimming of residual field harmonics using complex surface current-based coil patterns. Alternatively, one may modify and re-solve the governing equations to impose more complex primitives, such as combinations of axially symmetric and antisymmetric units with predetermined turn ratios. This may be useful for applications where a single coil is required to generate a target magnetic field profile composed of multiple field harmonics with different symmetries.

Our optimization approach may also be applied in other settings where there are diverse contributions that need to be minimized and maximized simultaneously, including aerofoil design~\cite{MUKESH2014191} and perturbation analysis in optimization problems~\cite{10.1007/978-1-4612-1394-9}.

\begin{appendices}

\section{Field harmonic components}\label{app.decomp}
Following Romeo~and~Hoult~\cite{Romeo}, we calculate the vector gradient of the spherical harmonics of degree $m\geq0$, \eqref{eq.sph}, to find the field harmonics, $\mathbf{B}_{n,m}\left(\mathbf{r}\right)=X_{n,m}\left(\mathbf{r}\right)\boldsymbol{\hat{x}} + Y_{n,m}\left(\mathbf{r}\right)\boldsymbol{\hat{y}} + Z_{n,m}\left(\mathbf{r}\right)\boldsymbol{\hat{z}}$, where
\begin{align}\label{eq.shbx}
    X_{n,m}\left(r,\theta,\phi\right)&= \eta_{n,m}r^{n-1} \Big[ -\frac{1+\delta_{m,0}}{2} \ \times
    \nonumber \\
    &\hspace{-10pt} P_{n-1,m+1}\left(\cos\theta\right) \cos\left((m+1)\phi\right)
    \ + \nonumber \\
    & \hspace{0pt} \ \frac{\left(1-\delta_{m,0}\right)\left(n+m-1\right)\left(n+m\right)}{2}
    \ \times \nonumber \\
    & \hspace{10pt}P_{n-1,m-1}\left(\cos\theta\right)
    \cos\left((m-1)\phi\right)\Big],
\end{align}
\begin{align}\label{eq.shby}
    Y_{n,m}\left(r,\theta,\phi\right)&= \eta_{n,m}r^{n-1} \Big[ -\frac{1+\delta_{m,0}}{2} \ \times
    \nonumber \\
    &\hspace{-10pt} P_{n-1,m+1}\left(\cos\theta\right) \sin\left((m+1)\phi\right)
    \ - \nonumber \\
    & \hspace{0pt} \ \frac{\left(1-\delta_{m,0}\right)\left(n+m-1\right)\left(n+m\right)}{2}
    \ \times \nonumber \\
    & \hspace{10pt}P_{n-1,m-1}\left(\cos\theta\right)
    \sin\left((m-1)\phi\right)\Big],
\end{align}
\begin{align}\label{eq.shbz}
    Z_{n,m}\left(r,\theta,\phi\right)= \eta_{n,m}& \left(n+m\right)r^{n-1} \ \times \nonumber \\
    &\hspace{10pt} P_{n-1,m}\left(\cos\theta\right)
    \cos\left(m\phi\right).
\end{align}

\section{Matching cylindrical and spherical field variations}\label{app.match}
Firstly, let us consider the standard series expansions~\cite{mathsbook},
\begin{equation}\label{eq.simple_taylor1}
    I_{m}(x)=\sum_{l=0}^\infty\ \frac{1}{l!(l+m)!}\left(\frac{x}{2}\right)^{2l+m},
\end{equation}
\begin{equation}\label{eq.simple_taylor2}
    \cos(y)=\sum_{l=0}^\infty\ \frac{(-1)^ly^{2l}}{(2l)!}.
\end{equation}
To match the variations in equations~\eqref{eq.simple_taylor1}~and~\eqref{eq.simple_taylor2} with the arguments of equation~\eqref{eq.bzdis+}, $x=k\rho/\rho_c$ and $y=kz/\rho_c$, we group the terms with common exponents of
$k$, and then convert to spherical coordinates, $\rho=r\sin\theta$ and $z=r\cos\theta$. We find
\begin{align}
 I_{m}\left(\frac{k\rho}{\rho_c}\right) \cos\left(\frac{kz}{\rho_c}\right) &= \nonumber \\ &\hspace{-70pt}\left(\frac{kr\sin\theta}{2\rho_c}\right)^m \sum^{\infty}_{{\nu}=0} \sum^{{\nu}}_{l=0}\ \frac{1}{({\nu}-l)!({\nu}-l+m)!} \ \times \nonumber \\ &\hspace{-60pt} \frac{(-1)^l}{2^{2({\nu}-l)}(2l)!} \left(\frac{kr}{\rho_c}\right)^{2{\nu}} \cos^{2l}\theta \sin^{2({\nu}-l)}\theta. \label{eq:Imcos_taylor}
\end{align}
We now introduce the expansion of the associated Legendre polynomials, as derived in reference~\cite{10.1021/ed077p244},
\begin{align}\label{eq.Pnm_taylor}
    P_{n,m}(x) &= 2^{n} (1-x^2)^{m/2} \ \times \nonumber \\  &\hspace{20pt} \sum_{l=m}^n\ \frac{l!}{(l-m)!} x^{l-m} \binom{n}{l} \binom{\frac{n+l-1}{2}}{n}.
\end{align}
We substitute $x=\cos\theta$ into equation~\eqref{eq.Pnm_taylor} and re-index to $n=2\nu+m$ and $l=l'+m$, finding
\begin{align}
    P_{2{\nu}+m,m}(\cos \theta)&= 2^{2{\nu}+m} \sin^m\theta \ \times \nonumber \\  &\hspace{-60pt} \sum_{l'=0}^{2{\nu}}\ \frac{(l'+m)!}{l'!} \binom{2{\nu}+m}{l'+m} \binom{{\nu}+m+\frac{l'-1}{2}}{2{\nu}+m} \cos^{l'}\theta. \label{eq:Pnm_taylor2}
\end{align}
For a fixed value of $\nu$, one may factorize out the variations proportional to $\sin^{m}\theta$ in equations~\eqref{eq:Imcos_taylor}~and~\eqref{eq:Pnm_taylor2} by examining the $l=l'=0$ terms. We can therefore match these variations together, with an unknown scaling, and invert to find the scaling. We find
\begin{align}
     I_{m}\left(\frac{k\rho}{\rho_c}\right) \cos\left(\frac{kz}{\rho_c}\right) &= \sum^{\infty}_{{\nu}=0}\ \frac{(-1)^{{\nu}}}{(2({\nu}+m))!}  \ \times \nonumber \\ & \left(\frac{kr}{\rho_c}\right)^{2{\nu}+m} P_{2{\nu}+m,m}\left(\cos \theta \right).
\end{align}
Similarly, following the same method using the standard series expansion of $\sin(y)$~\cite{mathsbook} and re-indexing equation~\eqref{eq.Pnm_taylor} to $n=2\nu+m-1$, we find
\begin{align}
    I_{m}\left(\frac{k\rho}{\rho_c}\right) \sin\left(\frac{kz}{\rho_c}\right) &= \sum^{\infty}_{{\nu}=1}\ \frac{(-1)^{{\nu}}}{(2({\nu}+m)-1)!} \ \times \nonumber \\ &\hspace{-20pt} \left(\frac{kr}{\rho_c}\right)^{2{\nu}+m-1} P_{2{\nu}+m-1,m}\left(\cos \theta \right).
\end{align}

\section{Solving the loops and saddles harmonic weighting function}\label{app.looparcint}
Here, we obtain the complete class of integrals
\begin{align}
\beta^+_{2{\nu}+m+1,m}\left(\chi_c\right) &= \nonumber \\ &\hspace{-40pt} (-1)^{{\nu+1}}\int_{0}^{\infty} \mathrm{d}k \ k^{2{\nu}+m+1} \cos(k\chi_c) K'_{m}(k), \label{eq:B^m_sym}
\end{align} 
\begin{align}
\beta^-_{2{\nu}+m,m}\left(\chi_c\right) &= \nonumber \\ &\hspace{-40pt} (-1)^{{\nu+1}}\int_{0}^{\infty} \mathrm{d}k \ k^{2{\nu}+m} \sin(k\chi_c) K'_{m}(k), \label{eq:B^m_anti}
\end{align} 
which we shall refer to as the symmetric and antisymmetric cases, respectively, for $\nu\in\mathbb{Z}^{0+}$ and $m\in\mathbb{Z}^{0+}$. We shall use the standard result~\cite{gradshteyn2007}
\begin{align}
\int_{0}^{\infty} \mathrm{d}k \ k^m \cos(k\chi_c)  K_{m}(k) &= \nonumber \\ &\hspace{-20pt} \frac{{\pi}(2m)!}{2^{m+1}m!}\left(\frac{1}{1+\chi_c^2}\right)^{m+1/2}. \label{eq:B^m_stan}
\end{align}

First, we integrate the left-hand-side of equation~\eqref{eq:B^m_stan} by parts,
\begin{align}
\int_{0}^{\infty} \mathrm{d}k \ k^m \cos(k\chi_c) K_{m}(k) &= \nonumber \\ &\hspace{-80pt}-\frac{1}{\chi_c}\int_{0}^{\infty} \mathrm{d}k \ k^{m}\sin(k\chi_c) K'_{m}(k) \ \nonumber \\ &\hspace{-60pt} -\frac{m}{\chi_c}\int_{0}^{\infty} \mathrm{d}k \ k^{m-1}\sin(k\chi_c) K_{m}(k),
\end{align}
which we can rearrange to show
\begin{align}\label{eq:B^mp}
\int_{0}^{\infty} \mathrm{d}k \ k^{m}\sin(k\chi_c) K'_{m}(k) &= \nonumber \\ &\hspace{-70pt} -\chi_c\int_{0}^{\infty} \mathrm{d}k \ k^m \cos(k\chi_c) K_{m}(k) \ \nonumber \\ &\hspace{-50pt} -m\int_{0}^{\infty} \mathrm{d}k \ k^{m-1}\sin(k\chi_c) K_{m}(k).
\end{align}
We can substitute equation~\eqref{eq:B^m_stan} into this to find
\begin{align}\label{eq:B_step}
\int_{0}^{\infty} \mathrm{d}k \ k^{m}\sin(k\chi_c) K'_{m}(k) &= \nonumber \\ &\hspace{-70pt} -\frac{{\pi}{\chi_c}(2m)!}{2^{m+1}m!}\left(\frac{1}{1+\chi_c^2}\right)^{m+1/2} \  \nonumber \\ &\hspace{-50pt} -m\int_{0}^{\infty} \mathrm{d}k \ k^{m-1}\sin(k\chi_c) K_{m}(k).
\end{align}
Now, we note that the derivative of the second term on the right-hand-side of equation~\eqref{eq:B_step} with respect to $\chi_c$ relates to equation~\eqref{eq:B^m_stan} via
\begin{align}
\frac{\partial}{\partial\chi_c}\left(\int_{0}^{\infty} \mathrm{d}k \ k^{m-1}\sin(k\chi_c) K_{m}(k) \right) &= \nonumber \\ &\hspace{-70pt} \int_{0}^{\infty} \mathrm{d}k \ k^m \cos(k\chi_c)  K_{m}(k),
\end{align}
and so
\begin{align}\label{eq:B^m_mini}
\int_{0}^{\infty} \mathrm{d}k \ k^{m-1}\sin(k\chi_c) K_{m}(k) &= \nonumber \\ &\hspace{-80pt} \int\mathrm{d}\chi_c\ \left( \frac{ \pi(2m)!}{2^{m+1}m!}\left(\frac{1}{1+\chi_c^2}\right)^{m+1/2} \right).
\end{align}
To solve equation~\eqref{eq:B^m_mini} we perform the substitution $\chi_c=\tan\theta_c$, and so $\cos\theta_c=1/\sqrt{1+\chi_c^2}$, finding
\begin{equation}\label{eq:B^m_mini2}
\int \mathrm{d}\chi_c\ \left(\frac{1}{1+\chi_c^2}\right)^{m+1/2} = \int\mathrm{d}\theta_c\ \cos^{2m-1} \theta_c.
\end{equation}
Changing variables to $u=\sin\theta_c$, equation~\eqref{eq:B^m_mini2} can be expanded as
\begin{align}\label{eq:B^m_mini3}
 \int\mathrm{d}\theta_c\ \cos^{2m-1} \theta_c &= \int\mathrm{d}u\ \left(1-u^2\right)^{m-1} \nonumber \\
 &= \sum_{k=0}^{m-1} \frac{(-1)^{k}}{2k+1} \binom{m-1}{k} u^{2k+1},
\end{align}
where we have applied the binomial theorem,
\begin{equation}
(A+B)^n = \sum_{k=0}^n \binom{n}{k} A^{n-k}B^k.
\end{equation}
Therefore, equation~\eqref{eq:B^m_mini} can be expressed as
\begin{align}
\int_{0}^{\infty} \mathrm{d}k \ k^{m-1}\sin(k\chi_c) K_{m}(k) &= \hspace{40pt} \nonumber \\ &\hspace{-115pt} \frac{ \pi(2m)!}{2^{m+1}m!} \sum_{k=0}^{m-1} \frac{(-1)^{k}}{2k+1} \binom{m-1}{k} \left(\frac{\chi_c^2}{1+\chi_c^2}\right)^{k+1/2}.
\end{align} 

Then, we note that we can perform derivatives of equation~\eqref{eq:B^mp} with respect to $\chi_c$ an odd or even number of times to match equations~\eqref{eq:B^m_sym}~and~\eqref{eq:B^m_anti},
\begin{align}
\frac{\partial^{2{\nu}+1}}{\partial\chi_c^{2{\nu}+1}}\left(\int_{0}^{\infty} \mathrm{d}k \ k^m \sin(k\chi_c) K'_{m}(k) \right) &= (-1)^{{\nu}} \ \times \nonumber \\ &\hspace{-100pt} \int_{0}^{\infty} \mathrm{d}k \ k^{2{\nu}+m+1} \cos(k\chi_c) K'_{m}(k),
\end{align}
\begin{align}
\frac{\partial^{2{\nu}}}{\partial\chi_c^{2{\nu}}}\left(\int_{0}^{\infty} \mathrm{d}k \ k^m \sin(k\chi_c) K'_{m}(k) \right) &= (-1)^{{\nu}} \ \times \nonumber \\ &\hspace{-100pt} \int_{0}^{\infty} \mathrm{d}k \ k^{2{\nu}+m} \sin(k\chi_c) K'_{m}(k).
\end{align}
Thus, the symmetric and antisymmetric integral cases, \eqref{eq:B^m_sym}~and~\eqref{eq:B^m_anti}, have the same result,
\begin{align}
\beta_{n+m,m}\left(\chi_c\right) &=  \frac{{\pi}(2m)!}{2^{m+1}m!} \frac{\partial^{n}}{\partial\chi_c^{n}} \Bigg[ \chi_c\left(\frac{1}{1+\chi_c^2}\right)^{m+1/2} \hspace{120pt} \nonumber \\ &\hspace{-25pt} +m \sum_{k=0}^{m-1} \frac{(-1)^{k}}{2k+1} \binom{m-1}{k}\left(\frac{\chi_c^2}{1+\chi_c^2}\right)^{k+1/2} \Bigg], \label{eq:B^m_final}
\end{align}
where $n=2\nu+1$ in the symmetric case and $n=2\nu$ in the antisymmetric case. In the case where $m=0$, equation~\eqref{eq:B^m_final} simplifies to
\begin{equation}
\beta_{n,0}\left(\chi_c\right) = \frac{\pi}{2} \frac{\partial^{n}}{\partial\chi_c^{n}}\left(\frac{\chi_c}{\sqrt{1+\chi_c^2}}\right). \label{eq:B^m_0_final}
\end{equation} 

\section{Solving the ellipses harmonic weighting function}\label{app.ellipseint}
Here, we obtain the complete class of integrals
\begin{align}
\gamma^-_{2(\nu+\mu),2\mu}\left(\chi_c,\psi_c\right) &= (-1)^{\nu+\mu} \times \nonumber \\ &\hspace{-50pt} \int_{0}^{\infty} \mathrm{d}k \ k^{2(\nu+\mu)} J_{2\mu}\left(k\psi_c\right)  \sin(k\chi_c) K'_{2\mu}(k),
\end{align} 
\begin{align}
\gamma^-_{2(\nu+\mu)-1,2\mu-1}\left(\chi_c,\psi_c\right) &= (-1)^{\nu+\mu} \times \hspace{120pt} \nonumber \\ &\hspace{-85pt} \int_{0}^{\infty} \mathrm{d}k \ k^{2(\nu+\mu)-1} J_{2\mu-1}\left(k\psi_c\right)  \cos(k\chi_c) K'_{2\mu-1}(k),
\end{align}
which we refer to as the even and odd degree antisymmetric cases, respectively, and
\begin{align}
\gamma^+_{2({\nu}+\mu)+1,2\mu}\left(\chi_c,\psi_c\right)&=\frac{\partial\gamma^-_{2(\nu+\mu),2\mu}\left(\chi_c,\psi_c\right)}{\partial\chi_c},
\end{align}
\begin{align}
\gamma^+_{2(\nu+\mu),2\mu-1}\left(\chi_c,\psi_c\right)&=\frac{\partial\gamma^-_{2(\nu+\mu)-1,2\mu-1}\left(\chi_c,\psi_c\right)}{\partial\chi_c},
\end{align}
which we refer to as the even and odd degree symmetric cases, respectively, for $0<\psi_c<\chi_c$, $\nu\in\mathbb{Z}^{0+}$, and $\mu\in\mathbb{Z}^{+}$. Here, we define
\begin{equation}\label{eq:g_stan1}
S_m\left(\chi_c,\psi_c\right) = i^m \tilde{S}_m\left(\chi_c,\psi_c\right)
\end{equation}
where
\begin{align}\label{eq:g_stan1.1}
\tilde{S}_m\left(\chi_c,\psi_c\right) = \int_{0}^{\infty} \mathrm{d}k \ J_{m}(k\psi_c) \cos(k\chi_c)  K_{m}(k).
\end{align}
Equation~\eqref{eq:g_stan1.1} has the known solution~\cite{gradshteyn2007}
\begin{align}
\tilde{S}_m\left(\chi_c,\psi_c\right) = \frac{(1-i)}{2i^m\sqrt{2\psi_c}} Q_{m-1/2}\left(i\frac{\psi_c^2-\chi_c^2-1}{2\psi_c}\right) \label{eq:g_stan2},
\end{align}
where $Q_{m-1/2}\left(z\right)$ is a Legendre polynomial of the second kind of half-integer order, $m-1/2$, for $m\in\mathbb{Z}^+$.

Now, let us integrate the right-hand-side of equation~\eqref{eq:g_stan1.1} by parts,
\begin{align}\label{eq:G^mp}
\int_{0}^{\infty} \mathrm{d}k \ J_{m}(k\psi_c) \cos(k\chi_c) k^{m} K_{m}(k) &= \nonumber  \hspace{100pt} \\ &\hspace{-140pt} -\frac{\psi_c}{\chi_c}\int_{0}^{\infty} \mathrm{d}k \ J_{m-1}(k\psi_c) \sin(k\chi_c) k^{m} K_{m}(k) \nonumber \\ &\hspace{-110pt} -\frac{1}{\chi_c}\int_{0}^{\infty} \mathrm{d}k \ J_{m}(k\psi_c) \sin(k\chi_c) k^{m} K'_{m}(k).
\end{align}
We shall use a standard formula for the derivative of the Bessel function of the first kind~\cite{mathsbook},
\begin{align}\label{eq:Jm_k}
J_{m-1}(k\psi_c) = \frac{1}{k}\left[\frac{\partial J_{m}(k\psi_c)}{\partial\psi_c} + \frac{m}{\psi_c}J_{m}(k\psi_c)\right].
\end{align}
We substitute equation~\eqref{eq:Jm_k} into equation~\eqref{eq:G^mp} and rearrange to find
\begin{align}\label{eq:G^mp2}
&\int_{0}^{\infty} \mathrm{d}k \ J_{m}(k\psi_c) \sin(k\chi_c) k^{m} K'_{m}(k) = \Bigg[ \hspace{160pt} \nonumber \\ &\hspace{-5pt}-\chi_c\int_{0}^{\infty} \mathrm{d}k \ J_{m}(k\psi_c) \cos(k\chi_c) k^{m} K_{m}(k) \nonumber \\ &\hspace{-5pt} -m\int_{0}^{\infty} \mathrm{d}k \ J_{m}(k\psi_c) \sin(k\chi_c) k^{m-1} K_{m}(k) \nonumber \\ &\hspace{-5pt} -\psi_c\frac{\partial}{\partial\psi_c}\Big(\int_{0}^{\infty} \mathrm{d}k \ J_{m}(k\psi_c) \sin(k\chi_c) k^{m-1} K_{m}(k)\Big) \Bigg].
\end{align}
Now, substituting $m=2\mu$ into equations~\eqref{eq:g_stan1}~and~\eqref{eq:g_stan1.1}, we can perform derivatives with respect to $\chi_c$, $2\mu$ and $\left(2\mu-1\right)$ times, to find
\begin{align}
 \frac{\partial^{2\mu}S_m\left(\chi_c,\psi_c\right)}{\partial\chi_c^{2\mu}} &= \nonumber \\ &\hspace{-25pt} \int_{0}^{\infty} \mathrm{d}k \ k^{2\mu} J_{2\mu}(k\psi_c) \cos(k\chi_c) K_{2\mu}(k), \label{eq:g_stan3}
\end{align}
and
\begin{align}
\frac{\partial^{2\mu-1}S_m\left(\chi_c,\psi_c\right)}{\partial\chi_c^{2\mu-1}} &= \nonumber \\ &\hspace{-35pt} \int_{0}^{\infty} \mathrm{d}k \ k^{2\mu-1} J_{2\mu}(k\psi_c) \sin(k\chi_c) K_{2\mu}(k), \label{eq:g_stan4}
\end{align}
respectively. Then, substituting equations~\eqref{eq:g_stan3}~and~\eqref{eq:g_stan4} into equation~\eqref{eq:G^mp2}, for $m=2\mu$, we find
\begin{align}\label{eq:G^m_inter1}
\int_{0}^{\infty} \mathrm{d}k \ J_{2\mu}(k\psi_c) \sin(k\chi_c) k^{2\mu} K'_{2\mu}(k) &= \Bigg[ \nonumber \hspace{160pt} \\ &\hspace{-150pt} -\chi_c \frac{\partial^{2\mu} S_{m}\left(\chi_c,\psi_c\right)}{\partial\chi_c^{2\mu}} -2\mu \frac{\partial^{2\mu-1} S_{m}\left(\chi_c,\psi_c\right)}{\partial\chi_c^{2\mu-1}} \nonumber \\ &\hspace{-30pt}  - \psi_c \frac{\partial^{2\mu} S_{m}\left(\chi_c,\psi_c\right)}{\partial\chi_c^{2\mu-1}\partial\psi_c}\Bigg].
\end{align}

We can follow the same steps as applied to equation~\eqref{eq:G^mp} with the integral
\begin{equation}
\int_{0}^{\infty} \mathrm{d}k \ J_{m}(k\psi_c) \sin(k\chi_c) k^{m} K_{m}(k),
\end{equation}
for $m=2\mu-1$, to find
\begin{align}\label{eq:G^m_inter2}
\int_{0}^{\infty} \mathrm{d}k \ J_{2\mu-1}(k\psi_c) \cos(k\chi_c) k^{2\mu-1} K'_{2\mu-1}(k) &= i\Bigg[ \nonumber \hspace{100pt} \\ &\hspace{-180pt} -\chi_c \frac{\partial^{2\mu-1} S_{m}\left(\chi_c,\psi_c\right)}{\partial\chi_c^{2\mu-1}} -\left(2\mu-1\right) \frac{\partial^{2(\mu-1)} S_{m}\left(\chi_c,\psi_c\right)}{\partial\chi_c^{2(\mu-1)}} \nonumber \\ &\hspace{-90pt}  - \psi_c \frac{\partial^{2\mu-1} S_{m}\left(\chi_c,\psi_c\right)}{\partial\chi_c^{2(\mu-1)}\partial\psi_c}\Bigg].
\end{align}
Then, we can finally generalize all cases in equations~\eqref{eq:G^m_inter1}~and~\eqref{eq:G^m_inter2} where the degree is even or odd and perform derivatives with respect to $\chi_c$, to find
\begin{align}\label{eq:G^m_sol1}
\gamma_{n+m,m}\left(\chi_c,\psi_c\right) &= (-1)^{m+1}\frac{\partial^{n}}{\partial\chi_c^{n}}\Bigg[\chi_c \frac{\partial^{m} \tilde{S}_{m}\left(\chi_c,\psi_c\right)}{\partial\chi_c^{m}} \ + \hspace{200pt} \nonumber \\ &\hspace{-40pt} m \frac{\partial^{m-1} \tilde{S}_{m}\left(\chi_c,\psi_c\right)}{\partial\chi_c^{m-1}} + \psi_c \frac{\partial^{m} \tilde{S}_{m}\left(\chi_c,\psi_c\right)}{\partial\chi_c^{m-1}\partial\psi_c}\Bigg],
\end{align} 
where $n=2\nu+1$ in the symmetric cases and $n=2\nu$ in the antisymmetric cases.

\section{Supplementary coil designs}\label{app.designs}
\begin{figure}[!ht]
    \centering
    \begin{tabular}{c c}
    \raisebox{-.5\height}{\scalebox{0.7857}{\includegraphics[page=25]{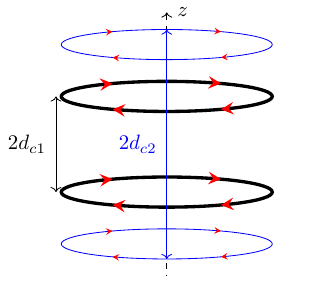}}} & \raisebox{-.5\height}{\scalebox{0.7857}{\includegraphics[page=26]{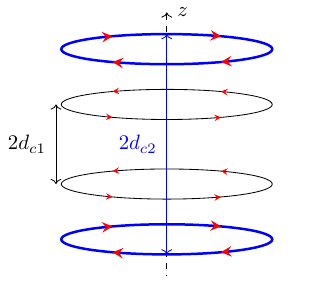}}} \\
    \small{(a)} & \small{(b)}
    \end{tabular}
    \caption{Schematics of $N_{\mathrm{loops}}$ axially symmetric loop pairs (red arrows indicate current flow direction) of radius $\rho_c$ at axial positions $z'={\pm}d_{ci}$ for $i\in\left[1:N_{\mathrm{loops}}\right]$. (a) $N_{\mathrm{loops}}=2$ pairs of loops with $d_{ci} =[0.4537, 0.9454]\rho_c$ and turn ratios $3:1$ (black, blue, respectively, with higher turn ratios in bold lines) (b) $N_{\mathrm{loops}}=2$ pairs of loops with $d_{ci} =[0.3775, 0.9020]\rho_c$ and turn ratios $-1:2$, labelled as (a).}
    \label{fig.optloopsbias}
\end{figure}%
\begin{figure}[!ht]
\begin{tabular}{c c}
\includegraphics[page=27]{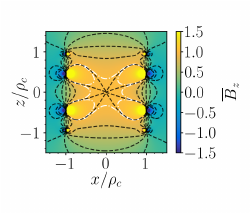} & \includegraphics[page=28]{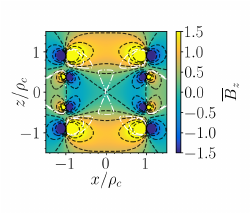} \\
\small{(a)}\hspace{20pt} & \small{(b)}\hspace{20pt}
\end{tabular}
\caption{Magnitude of the normalized axial magnetic field (color scales right), $\overline{B}_z=B_z/B_0$, where $B_0$ is the magnetic field gradient strength specified in Table~\ref{table.stats.bias}, in the $xz$-plane generated by the coils in Fig.~\ref{fig.optloopsbias}a and Fig.~\ref{fig.optloopsbias}b, corresponding to (a) and (b), respectively, where the coils are of radius $\rho_c$. White contours enclose the regions where (a) $B_z$ and (b) $\mathrm{d}^2B_z/\mathrm{d}z^2$ deviate from perfect uniformity and curvature, respectively, by less than $1$\% and $10$\%, respectively (dot-dashed curves).}
\label{fig.biascomp}
\end{figure}
\begin{table}[!ht]
\caption{Properties of the optimized coils in Fig.~\ref{fig.optloopsbias}, here given for a coil radius of $\rho_c=10$~cm and labelled as Table~\ref{table.stats.discrete}. The magnetic field strength, $B_0$, is the value of $B_z$ for $\mathbf{B}_{1,0}$ and $\mathrm{d}^2B_z/\mathrm{d}z^2$ for $\mathbf{B}_{3,0}$ and is evaluated at the centre of the coil per unit current, $I$.}
\label{table.stats.bias}
    \centering
    {\setlength{\arrayrulewidth}{0.25mm}
    \setlength{\tabcolsep}{2pt}
    \renewcommand{\arraystretch}{1.1}
    \begin{tabular}{|P{0.1\columnwidth} | P{0.14\columnwidth} || P{0.14\columnwidth} | P{0.14\columnwidth} | P{0.18\columnwidth} | P{0.14\columnwidth} |}
    \hline
    $\mathbf{B}_{n,m}$ & Coil &  $\mathrm{max}\left(\chi_c\right)$, (cm) & $l$, (m) & $B_0/{I}$ & $L$, ($\mu$H) \\
    \hline \hline
    $\mathbf{B}_{1,0}$ & Fig.~\ref{fig.optloopsbias}a & $9.45$ & $5.03$ & $33.3$~$\mu$T/A & $13.8$ \\
    \hline \hline
    $\mathbf{B}_{3,0}$ & Fig.~\ref{fig.optloopsbias}b & $9.02$ & $3.77$ & $3.2$~mT/Am\textsuperscript{2} & $4.69$ \\
    \hline
\end{tabular}}
\end{table}
Sets of symmetric loops are presented in Fig.~\ref{fig.optloopsbias} which are optimized to generate the $\mathbf{B}_{1,0}$ and $\mathbf{B}_{3,0}$ field harmonics, displayed in equations~\eqref{eq.B10}~and
\begin{align}
&\mathbf{B}_{3,0}\left(x,y,z\right) = \frac{3}{4}\sqrt{\frac{7}{\pi}}\ (-2xz\boldsymbol{\hat{x}}- \nonumber \\ &\hspace{100pt} 2yz\boldsymbol{\hat{y}}+\left(-x^2-y^2+2z^2\right)\boldsymbol{\hat{z}}). \label{eq.B30}
\end{align}
These designs are optimized using our \textit{Mathematica} program stored in the repository listed in reference~\cite{NoahRepo}. The magnetic fields generated by the coils are displayed in Fig.~\ref{fig.biascomp} and their properties are summarized in Table~\ref{table.stats.bias}.

\end{appendices}

\section*{Acknowledgements}
We acknowledge the support of the UK Quantum Technology Hub Sensors and Timing (EP/T001046/1) and from Innovate UK Project 44430 MAG-V: Enabling Volume Quantum Magnetometer Applications through Component Optimization \& System Miniaturization. \\

We also acknowledge Sionnach Devlin, a Senior Technician in the School of Physics \& Astronomy workshop, for his assistance 3D-printing the coil former.

\section*{Declarations}
The authors P.J.H and N.L.H contributed equally to this work. \\

P.J.H, M.P, M.B, R.B, and M.F have a worldwide patent (WO/2021/053356) which includes some of coil design techniques applied in this work. N.L.H is a paid employee of Wolfram Research Inc., who make \textit{Mathematica}. N.H, M.B, and R.B hold founding equity in Cerca Magnetics Limited, who commercialize atomic magnetometer technology. T.S, C.M, A.D, and M.A.W declare no competing interests. \\

All supporting data may be made available on request. Our \textit{Mathematica} program used to generate the coil designs presented in this work is publicly-available for non-commercial use~\cite{NoahRepo}. \\

This work has been submitted to the IEEE for possible publication. Copyright may be transferred without notice, after which this version may no longer be accessible.

\bibliographystyle{IEEEtran}
\bibliography{literatur}

\end{document}